\documentclass[12pt,english,floatfix,superscriptaddress,aps,prd,preprint,showkeys,nofootinbib]{revtex4-1}
\usepackage{amsmath}
\usepackage{amssymb}
\usepackage{amsbsy}
\usepackage{amsfonts}
\usepackage{amsopn}
\usepackage{amstext}
\usepackage{graphicx}
\usepackage[english]{babel}
\usepackage{color}
\usepackage{slashed}
\usepackage{esint}
\usepackage[dvips]{epsfig}
\usepackage[dvips]{graphicx}
\usepackage{float}
\usepackage{units}
\usepackage{textcomp}
\usepackage{wasysym}
\usepackage{hyperref}
\usepackage{slashed}
\usepackage[table,xcdraw]{xcolor}
\usepackage[linesnumbered,ruled,vlined]{algorithm2e}
\usepackage{tikz}
\usetikzlibrary{quantikz}
\usepackage{physics}

\begin{document}
\title{Parameter optimization comparison in QAOA between Stochastic Hill Climbing with Random Re-starts and Local Search with entangled and non-entangled mixing operators}
\author{Brian García Sarmina}
\email{brian.garsar.6@gmail.com}
\affiliation{Centro de Investigaci\'{o}n en Computaci\'{o}n, Instituto Polit\'{e}cnico Nacional, UPALM, CDMX 07700, Mexico.}
\author{Guo-Hua Sun}
\email{sunghdb@yahoo.com}
\affiliation{Centro de Investigaci\'{o}n en Computaci\'{o}n, Instituto Polit\'{e}cnico Nacional, UPALM, CDMX 07700, Mexico.}
\author{Shi-Hai Dong}
\email{dongsh2@yahoo.com}
\affiliation{Centro de Investigaci\'{o}n en Computaci\'{o}n, Instituto Polit\'{e}cnico Nacional, UPALM, CDMX 07700, Mexico.}
\affiliation{Reserach Center for Quantum Physics, Huzhou University, Huzhou, 313000, P. R. China.}


\begin{abstract}
This study investigates the efficacy of Stochastic Hill Climbing with Random Restarts (SHC-RR) and Local Search (LS) --using multiplication operation for the application of perturbation and LS* using summation operation for the application of perturbation-- strategies within the Quantum Approximate Optimization Algorithm (QAOA) framework across various problem models. Employing uniform parameter settings, including the number of restarts and SHC steps, we analyze LS with two distinct perturbation operations: multiplication and summation. Our comparative analysis encompasses multiple versions of max-cut and random Ising model (RI) problems, utilizing QAOA models with depths ranging from $1L$ to $3L$. These models incorporate diverse mixing operator configurations, which integrate $RX$ and $RY$ gates, and explore the effects of an entanglement stage within the mixing operator. Our results consistently show that SHC-RR outperforms LS approaches, showcasing superior efficacy despite its ostensibly simpler optimization mechanism. Furthermore, we observe that the inclusion of entanglement stages within mixing operators significantly impacts model performance, either enhancing or diminishing results depending on the specific problem context.
\end{abstract}
\maketitle

\section{Introduction}

The Quantum Approximate Optimization Algorithm (QAOA), introduced by Farhi et al. in 2014 \cite{R1}, has attracted considerable attention owing to its potential applications in near-term quantum computing devices. Notably, QAOA stands out for its low depth and noise resistance, making it one of the most promising algorithms for tackling optimization problems on Noisy Intermediate-Scale Quantum (NISQ) devices \cite{R5, R6}. The term "low depth" refers to the relatively modest number of qubits required to represent and compute a given problem, accompanied by a potentially low count of quantum gates needed to construct the quantum circuit, depending on the problem at hand. These characteristics render QAOA particularly resilient to noise, a formidable challenge for near-term quantum computing systems. Numerous studies have delved into the efficacy of QAOA across a spectrum of optimization problems, including Max-Cut, the traveling salesman problem, Ising model problems, and graph partitioning, among others \cite{R1, R5, R6, R2, R3, R4, R7, R8, R9}.

While QAOA has predominantly found applications in combinatorial optimization problems such as Max-Sat, Max-Cut, and Ising Model \cite{R5, R6, R4, R10, R29}, it's worth noting that these problems can be viewed as generalizations of specific challenges with broader implications. Indeed, they have significant applications in diverse fields including finance \cite{R11, R12} and even quantum machine learning \cite{R13, R14}.

Optimizing variational quantum algorithms, such as QAOA, presents a significant challenge due to the limited understanding of how the parameter search space correlates with the Hilbert space \cite{R16, R26, R27}. Consequently, there exists a considerable amount of missing information and hypotheses that could be explored to enhance parameter optimization strategies. Classical methods for optimizing QAOA parameters have encountered difficulties stemming from the barren plateaus problem \cite{R17, R27}. This issue arises when certain regions of the parameter search space exhibit vanishing gradients, posing challenges for classical optimization techniques like gradient descent to find optimal values. In contrast, quantum optimization methods have the potential to explore the entire search space simultaneously, which may offer more efficient optimization solutions in the presence of barren plateaus, contingent upon the chosen optimization approach.

In light of the challenges encountered in optimizing QAOA, extensive research has delved into various optimization methods, with a primary focus on two key objectives. Firstly, addressing issues such as the well-known barren plateaus problem and other complexities that may impede the search for optimal parameters. Several studies \cite{R19, R20, R21} have identified a correlation between vanishing gradients, influenced by the number of qubits and the layers (comprising phase and mixing operators) within the Variational Quantum Algorithm (VQA) model. Secondly, the aim is to enhance the efficiency of the optimization process. Numerous studies \cite{R5, R6, R2, R4, R21, R28} have explored diverse parameter optimization techniques in VQAs. Often, the most effective approaches stem from classical heuristics or involve specially designed heuristics integrating certain quantum mechanical phenomena. Notably, some popular optimization methods include Stochastic Hill Climbing or Stochastic Gradient Descent. In our study, we adopted a similar approach for the internal parameter search.

Despite the extensive research on optimization methods for VQAs like QAOA, there are still unexplored questions and comparisons yet to be made. To address this gap, we propose a comprehensive comparison between three heuristic approaches: SHC-RR, LS, where the LS approach is tested using two operations for the perturbation step: multiplication and summation (LS*) \cite{R15}. LS is considered a refined version of SHC-RR \cite{R22, R32}. Our study involved testing these three methods across multiple instances of max-cut and random Ising model (RI) problems, varying the number of nodes, configurations, and connections for the different problems. The comparison between these approaches is based on their relative efficiency in achieving optimal parameter optimization. All methods have the potential to reach the global optimum solution given sufficient iterations. However, LS and LS* assumes a specific structure in exploring the search space, while SHC-RR employs a random-jumping search strategy. In SHC-RR, we make jumps called random restarts, followed by smaller jumps around that point using SHC. This strategy is somewhat similar in LS (or LS*), but here, the jumps are constrained by a perturbation factor or the local area explored by the heuristic, which expands with the number of allowed iterations.

To enrich the research presented in this paper alongside various optimization approaches, we also tested different QAOA models with varying depths, both with and without an entanglement stage in the mixing operator. This analysis yields interesting results concerning the optimization approaches tested. 

\section{QAOA description}

The QAOA operates by applying a sequence of unitary operators known as the \textbf{phase operator} $U(H, \gamma)$ and the \textbf{mixing operator} $U(B, \beta)$. These operators are applied sequentially to evolve the state $|\psi\rangle$. Adjusting the parameters $\gamma$ and $\beta$ enhances the likelihood of measuring the state $|\psi_{\gamma \beta}\rangle$, which encodes the solution to the problem at hand
\begin{equation}
    |\psi_{\gamma \beta} \rangle = U(B, \beta_{p}) U(H, \gamma_{p})\cdots U(B, \beta_{1})U(H, \gamma_{1}) |\psi \rangle.
    \label{eq:general_qaoa}
\end{equation}

The phase operator $U(H, \gamma)$ encodes the problem's Hamiltonian $H$ into a valid quantum operator. This Hamiltonian is commonly derived from the cost function of the original problem. When the quantum circuit applies $U(H, \gamma)$, it evolves the state through $RZ$ rotations, considering the interactions between the nodes or particles.
\begin{equation}
    U(H, \gamma) = e^{-i \gamma H}
    \label{eq:phase_operator}
\end{equation}

The mixing operator $U(B, \beta)$ plays a crucial role in inducing constructive or destructive interference among states, thereby influencing the prevalence of certain states over others when combined with the phase operator. Traditionally, $B$ consists solely of $RX$ rotations. However, as highlighted by D. Koch \textit{et al.} (2020) \cite{R10}, relying solely on $RZ$ and $RX$ rotations may not adequately cover the entire Hilbert space of the state $|\psi \rangle$. This limitation can render certain states unreachable for QAOA. To address this challenge and expand the coverage of QAOA, it's recommended to apply consecutive phase and mixing operators. In some cases, modifying the mixing operator becomes necessary. The proposed mixing operator incorporates $RY$ rotations and introduces an entangled stage. Our aim is to access a broader range of states and achieve maximally entangled states, following a similar approach outlined in \cite{R10}.
\begin{equation}
    U(B, \beta) = e^{-i \beta B}
    \label{eq:mixing_operator}
\end{equation}


But how does QAOA actually work? The idea behind QAOA is to generate rotations using $RZ$ ($\sigma_{z}$), $RX$ ($\sigma_{x}$), and in our model also $RY$ ($\sigma_{y}$), according to the parameters $\gamma$ and $\beta_{1,2}$ ($\beta_{1}$ for the $RX$ and $\beta_{2}$ for the $RY$), with the aim of finding the optimal rotations that will give us the state (or states) that encode the solution to our problem. The way the parameters affect the rotations arises from:
\begin{equation}
    \left [ \sigma_{z}, \sigma_{x} \right ] \neq 0,
    \label{eq:commutator_sigZ_sigX}
\end{equation}
and for our mixing operator also:
\begin{equation}
    \left [ \sigma_{z}, \sigma_{y} \right ] \neq 0,
    \label{eq:commutator_sigZ_sigY}
\end{equation}
\begin{equation}
    \left [ \sigma_{x}, \sigma_{y} \right ] \neq 0,
    \label{eq:commutator_sigX_sigY}
\end{equation}


\textbf{Equations} (\ref{eq:commutator_sigZ_sigX}), {(\ref{eq:commutator_sigZ_sigY})}, and {(\ref{eq:commutator_sigX_sigY})} depict the non-commutative behavior between the rotations, signifying a competitive interaction between stages. This competition, or incompatibility, is precisely what enables QAOA to function effectively. By fine-tuning the parameters $\gamma$ and $\beta_{1,2}$, we can balance the influence of the phase operator and the mixing operator, leading to an interference pattern that amplifies the likelihood of measuring certain states while reducing the probability of measuring others. This interference pattern serves as the quantum counterpart of a classical optimization landscape.

Similar to classical optimization, where we seek the global minimum of a cost function, QAOA searches for the quantum state(s) representing the optimal solution(s) to our problem. The interference pattern generated by QAOA guides our exploration towards these quantum states, enhancing the likelihood of successful exploration. This approach facilitates the evaluation of all potential state combinations, enabling the generation of an approximate solution for that specific parameter set in a single iteration.

\section{Problem description}

The problems under analysis encompass various versions of Max-Cut and random Ising model (RIM) problems. Specifically, the Max-Cut problems include cyclic and full (complete) configurations. Max-Cut is a well-studied optimization problem renowned for its simplicity and capacity to exhibit intricate behavior with increasing graph size. In the Max-Cut problem, we are presented with an undirected graph comprising vertices and edges connecting them. The objective is to partition the vertices into two sets in a manner that maximizes the number of edges between the two sets, thereby maximizing the cut.

This problem finds application across diverse domains such as network analysis, image segmentation, and clustering. In our analysis, we explore graphs of varying sizes and endeavor to identify the optimal solution leveraging the QAOA algorithm.

\begin{figure}[ht]
\centering
\includegraphics[width=8cm, height=6cm]{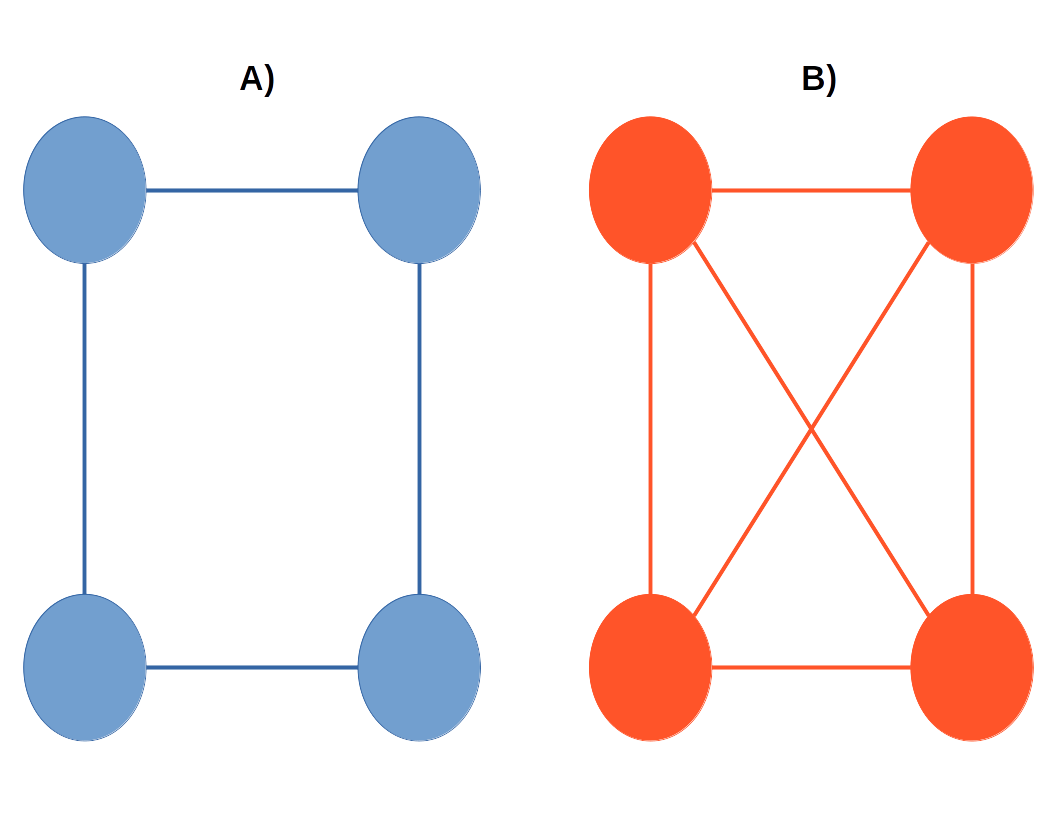}
\caption{General problem description for Max-Cut configurations: A) cyclic configuration 4 nodes and B) full (complete) configuration 4 nodes.}
\label{fig:gen_prob_description}
\end{figure}


In \textbf{Figure \ref{fig:gen_prob_description}}, we illustrate the general configurations of Max-Cut. Both the cyclic and full configurations will be investigated, with the number of nodes varying across 4, 10, and 15. In our full (or complete) configuration, every pair of nodes in the graph is connected by an edge.

For the Ising Model problems, our focus was on a variant called the random Ising model (RIM), which bears similarities to the max-cut problem. However, in the RIM, each particle (or node) is associated with its own magnetic field. We compiled a dataset comprising 100 RIM problems, with the number of particles ranging from 5 to 15. Additionally, the individual magnetic fields were randomly sampled from a uniform distribution spanning from -1 to 1.

\begin{figure}[ht]
\centering
\includegraphics[width=8cm, height=6cm]{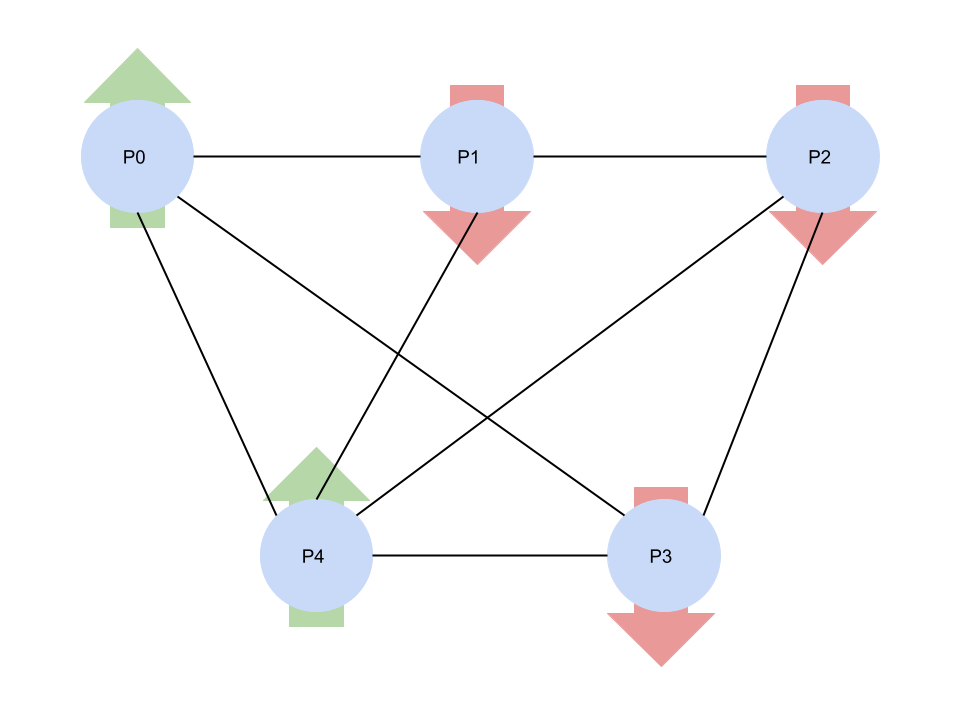}
\caption{Example of random Ising model problem.}
\label{fig:ex_random_ising_model}
\end{figure}

In Figure \ref{fig:ex_random_ising_model}, we present an example problem drawn from our generated set of RIM problems. These problems adhere to specific conditions, with a primary requirement being the presence of at least one connection between each pair of particles in the graph, akin to the max-cut cyclic configuration. In this example, the primary connections span from $P0$ to $P4$. Subsequently, we developed a subroutine to randomly establish connections between the nodes in the graph while ensuring no duplicate connections, as seen between $P1$ and $P4$, $P2$ and $P4$, and $P0$ and $P3$. Finally, individual magnetic fields are assigned random values ranging between -1 and 1, depicted by green arrows for positive magnetic fields and red arrows for negative magnetic fields in the provided illustration.

\subsection{Phase operator}

The equation of the general phase operator for Max-Cut problems can be expressed as

\begin{equation}
    U(H_{P}, \gamma) = e^{-i\gamma H_{P}} = \prod_{\langle j, k \rangle}^{} e^{-i\gamma Z_{j}Z_{k}}
    \label{eq:phase_operator_problem_cyclic}
\end{equation}

where $H$ represents the Hamiltonian provided in \textbf{(\ref{eq:phase_operator_problem_cyclic})} derived from the Max-Cut problem, and $\gamma$ is the parameter controlling the evolution of the quantum state through the phase rotations. This operator applies phase rotations to the quantum state, influenced by the problem's Hamiltonian, in order to encode information about the optimization landscape into the quantum state. In this equation, the notation $\langle j,k \rangle$ is used to denote the connections between neighboring nodes, where each node (or particle) is connected to its adjacent nodes, effectively forming a cycle. This arrangement includes two connections between each pair of adjacent nodes, and it completes the cycle by including an interaction (connection) between the first and last nodes.

\begin{equation}
   U(H_{P}, \gamma) = e^{-i\gamma H_{P}} = \prod_{ \left \{ j, k \mid j \neq k \right \} }^{} e^{-i\gamma Z_{j}Z_{k}}
    \label{eq:phase_operator_problem_complete}
\end{equation}

For the complete configuration case, we represent it with Eq. (\ref{eq:phase_operator_problem_complete}). In contrast to the cyclic configuration, the complete configuration establishes connections between every pair of nodes (particles) in the graph while excluding repeated connections and self-connections. This is denoted using the notation $\left\{ j, k \mid j \neq k \right\}$.

In the RIM problems, the phase operator could be represented in the following way:

\begin{equation}
    U(H_{P}, \gamma) = e^{-i\gamma H_{P}} = \prod_{\langle j, k \rangle}^{} e^{-i\gamma Z_{j}Z_{k}} \prod_{ j }^{} e^{-i \gamma h_{j} Z_{j}},\label{eq:phase_operator_RIM}
\end{equation}
where the term $\langle j,k \rangle$ encapsulates all connections within the specific RIM problem, and $h_{j}$ represents the associated individual magnetic field. Typically, the second term, corresponding to the magnetic field, is also influenced by the $\gamma$ parameter. Alternative approaches may introduce a new parameter associated with the magnetic field. However, such a method could potentially alter the problem's dynamics, as the relationship between the connections and the magnetic field might vary with different parameters.

\begin{figure}[ht]
    \begin{center}
        \begin{tikzpicture}
            \node[scale=0.65] {
                \begin{quantikz}
                    \ket{0}_{0}  & \gate{H} & \ctrl{1} &\qw & \ctrl{1} & \qw & \qw & \qw & \qw & \qw & \qw & \qw & \targ{} & \gate{RZ(-i\gamma)} & \targ{} & \qw \\
                    \ket{0}_{1} & \gate{H} & \targ{} & \gate{RZ(-i\gamma)} & \targ{} & \ctrl{1} & \qw & \ctrl{1} & \qw & \qw & \qw & \qw & \qw & \qw & \qw & \qw \\
                    \ket{0}_{2} & \gate{H} & \qw & \qw & \qw & \targ{} & \gate{RZ(-i\gamma)} & \targ{} & \qw & \ctrl{1} & \qw & \ctrl{1} & \qw & \qw & \qw & \qw \\
                    \ket{0}_{3} & \gate{H} & \qw & \qw & \qw & \qw & \qw & \qw & \qw & \targ{} & \gate{RZ(-i\gamma)} & \targ{} & \ctrl{-3} & \qw & \ctrl{-3} & \qw
                \end{quantikz}
            };
        \end{tikzpicture}
    \end{center}
    \caption{Quantum circuit for phase operator for cyclic configuration example.}
    \label{dia:phase_operator_cyclic_example}
\end{figure}
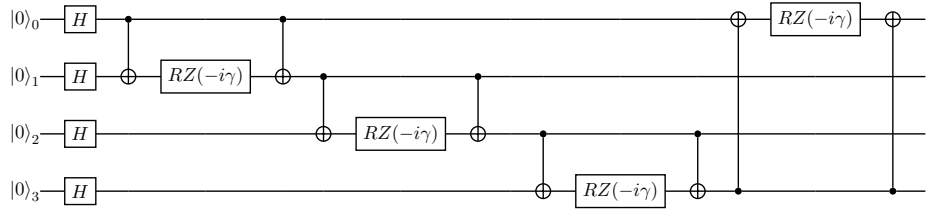

The circuit representation of the phase operator for the cyclic configuration for the example problem can be viewed in the \textbf{Figure \ref{dia:phase_operator_cyclic_example}}.

\begin{figure}[ht!]
    \begin{center}
        \begin{tikzpicture}
            \node[scale=0.65] {
                \begin{quantikz}
                    \cdots & & \qw & \qw & \qw & \qw & \qw & \qw & \qw \\
                    \cdots & & \ctrl{2} & \qw & \ctrl{2} & \qw & \qw & \qw & \qw \\
                    \cdots & & \qw & \qw & \qw & \ctrl{1} & \qw  & \ctrl{1} & \qw \\
                    \cdots & & \targ{} & \gate{RZ(-i\gamma)} & \targ{} & \targ{} & \gate{RZ(-i\gamma)} & \targ{} & \qw 
                \end{quantikz}
            };
        \end{tikzpicture}
    \end{center}
    \caption{Quantum circuit for phase operator for complete configuration example.}
    \label{dia:phase_operator_complete_example}
\end{figure}
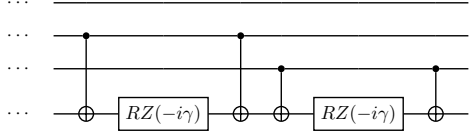

Continuing with the phase operator for the complete configuration for the example problem, we present the \textbf{Figure \ref{dia:phase_operator_complete_example}}.

\begin{figure}[ht!]
    \begin{center}
        \begin{tikzpicture}
            \node[scale=0.65] {
                \begin{quantikz}
                    \ket{0}_{0}  & \gate{H} & \ctrl{1} &\qw & \ctrl{1} & \ctrl{2} & \qw & \ctrl{2} & \ctrl{3} & \qw & \ctrl{3} & \qw & \qw & \qw & \cdots \\
                    \ket{0}_{1} & \gate{H} & \targ{} & \gate{RZ(-i\gamma)} & \targ{} & \qw & \qw & \qw & \qw & \qw & \qw & \ctrl{1} & \qw & \ctrl{1} & \cdots \\
                    \ket{0}_{2} & \gate{H} & \qw & \qw & \qw & \targ{} & \gate{RZ(-i\gamma)} & \targ{} & \qw & \qw & \qw & \targ{} & \gate{RZ(-i\gamma)} & \targ{} & \cdots \\
                    \ket{0}_{3} & \gate{H} & \qw & \qw & \qw & \qw & \qw & \qw & \targ{} & \gate{RZ(-i\gamma)} & \targ{} & \qw & \qw & \qw & \cdots
                \end{quantikz}
            };
        \end{tikzpicture}
    \end{center}
\end{figure}

Finally, for the case of the RIM problems, the quantum circuit representation adds an $RZ$ term related to the magnetic field $h_{j}$.

\begin{figure}[ht]
    \begin{center}
        \begin{tikzpicture}
            \node[scale=0.65] {
                \begin{quantikz}
                    \ket{0}_{0}  & \gate{H} & \ctrl{1} &\qw & \ctrl{1} & \qw & \qw & \qw & \qw & \ \ldots & & \gate{RZ(-i h_{0} \gamma)} & \qw \\
                    \ket{0}_{1} & \gate{H} & \targ{} & \gate{RZ(-i\gamma)} & \targ{} & \ctrl{1} & \qw & \ctrl{1} & \qw & \ \ldots & & \gate{RZ(-i h_{1} \gamma)} & \qw \\
                    \ket{0}_{2} & \gate{H} & \qw & \qw & \qw & \targ{} & \gate{RZ(-i\gamma)} & \targ{} & \qw & \ \ldots & & \gate{RZ(-i h_{2} \gamma)} & \qw \\
                    \ket{0}_{3} & \gate{H} & \qw & \qw & \qw & \qw & \qw & \qw & \qw &\ \ldots & & \gate{RZ(-i h_{3} \gamma)} & \qw \\
                \end{quantikz}
            };
        \end{tikzpicture}
    \end{center}
    \caption{Quantum circuit for phase operator for RIM problems.}
    \label{dia:phase_operator_RIM}
\end{figure}
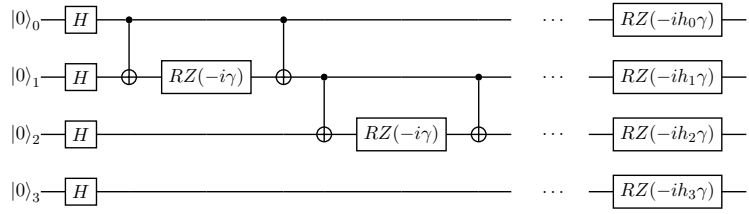

As depicted in \textbf{Figure \ref{dia:phase_operator_RIM}}, the interactions (connections) between qubits resemble those of the linear configuration, but we include additional connections that define a specific RIM problem (like those in \textbf{Figure \ref{fig:ex_random_ising_model}}). Finally, we incorporate the magnetic field component using $RZ$ gates with their corresponding field value $h_{j}$. This value may be positive or negative based on the field value. 

\subsection{Mixing operator}

For the mixing operator, we will present two general forms, inspired by the ideas proposed by D. Koch \textit{et al.} (2020) \cite{R10}. The first mixing operator will exclusively involve individual rotations using $RX$ and $RY$ gates. The second form of the mixing operator will incorporate an entanglement stage between the $RX$ and $RY$ gates.

The general equation for the mixing operator without an entanglement stage is the following:
\begin{equation}
    U(H_{B}, \beta_{n}) = e^{-i\beta_{n} H_{B}} = \prod_{ j }^{} e^{-i\beta_{1} X_{j}} e^{-i\beta_{2} Y_{j}} ,
    \label{eq:mixing_operator_cyclic_problem}
\end{equation}
from Eq.\textbf{(\ref{eq:mixing_operator_cyclic_problem})} it can be established that the mixing operator presents not only rotations along the $X$ axis but also on the $Y$ axis.

The mixing operator with an entanglement stage is represented by the following equation:
\begin{equation}
    U(H_{B}, \beta_{n}) = e^{i \beta_{n} H_{B}} = \prod_{ j }^{} e^{i\beta_{1} X_{j}} \prod_{ \left \{ j, k \mid j \neq k \right \} }^{} e^{i I_{j}X_{k}} \prod_{ j }^{}e^{i\beta_{2} Y_{j}}.
    \label{eq:mixing_operator_complete_problem}
\end{equation}
In this equation, the expansion of the CNOT gate within the term $\prod_{ \left \{ j, k \mid j \neq k \right \} }^{} e^{i I_{j}X_{k}}$. This term establishes control with $I_{j}$ and designates the target as $X_{k}$. By utilizing the same expression, we create a complete CNOT interaction between each qubit in the circuit, ensuring that repeated interactions and self-interactions are excluded from the entanglement stage in the mixing operator.

\begin{figure}[ht!]
    \begin{center}
        \begin{tikzpicture}
            \node[scale=0.65] {
                \begin{quantikz}
                    \ket{0}_{0}  & \gate{RX(-i\beta_{1})} & \qw & \gate{RY(-i\beta_{2})} & \qw \\
                    \ket{0}_{1} & \gate{RX(-i\beta_{1})} & \qw & \gate{RY(-i\beta_{2})} & \qw \\
                    \ket{0}_{2} & \gate{RX(-i\beta_{1})} & \qw & \gate{RY(-i\beta_{2})} & \qw \\
                    \ket{0}_{3} & \gate{RX(-i\beta_{1})} & \qw & \gate{RY(-i\beta_{2})} & \qw
                \end{quantikz}
            };
        \end{tikzpicture}
    \end{center}
    \caption{Quantum circuit for mixing operator without entanglement stage using problem example.}
    \label{dia:mixing_operator_example}
\end{figure}
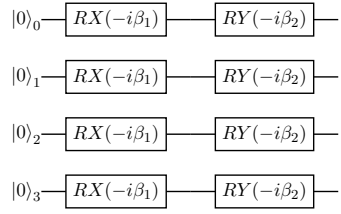

Translating the mixing operator equation without an entanglement stage into a quantum circuit representation, we arrive at the following diagram (\textbf{Diagram \ref{dia:mixing_operator_example}}) expression.

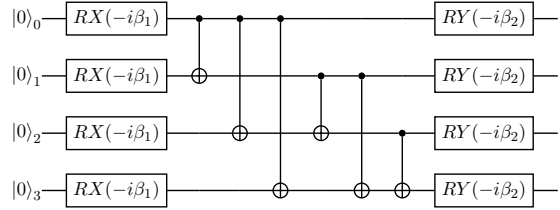
\begin{figure}[ht!]
    \begin{center}
        \begin{tikzpicture}
            \node[scale=0.65] {
                \begin{quantikz}
                    \ket{0}_{0}  & \gate{RX(-i\beta_{1})} & \ctrl{1} & \ctrl{2} & \ctrl{3} & \qw & \qw & \qw & \gate{RY(-i\beta_{2})} & \qw \\
                    \ket{0}_{1} & \gate{RX(-i\beta_{1})} & \targ{} & \qw & \qw & \ctrl{1} & \ctrl{2} & \qw & \gate{RY(-i\beta_{2})} & \qw \\
                    \ket{0}_{2} & \gate{RX(-i\beta_{1})} & \qw & \targ{} & \qw & \targ{} & \qw & \ctrl{1} & \gate{RY(-i\beta_{2})} & \qw \\
                    \ket{0}_{3} & \gate{RX(-i\beta_{1})} & \qw & \qw & \targ{} & \qw & \targ{} & \targ{} & \gate{RY(-i\beta_{2})} & \qw
                \end{quantikz}
            };
        \end{tikzpicture}
    \end{center}
    \caption{Quantum circuit for mixing operator with entanglement stage using problem example.}
    \label{dia:mixing_operator_example_ent}
\end{figure}

The second form of the mixing operator, which includes the entanglement stage, is depicted in the \textbf{Diagram \ref{dia:mixing_operator_example_ent}}.

As illustrated in \textbf{Figure \ref{dia:mixing_operator_example_ent}}, the entanglement state arises from a sequence of CNOT gates, generating a comprehensive configuration reminiscent of the complete max-cut problem. In this configuration, every qubit is intricately linked to each other without any redundant connections. While several techniques exist for establishing this entanglement stage, our study refrains from delving into the comparison of these methods and their effectiveness.

\section{Heuristics representation}

The optimization techniques employed in this paper encompass LS and SHC-RR. In this section, we elucidate the formulations of these methodologies for implementation. It's noteworthy that our heuristic model of LS can be construed as an enhanced iteration of SHC-RR. The local search methodology endeavors to anticipate or project a particular behavior within the search space, thereby facilitating more efficient exploration. Conversely, the SHC-RR heuristic doesn't predetermine a specific behavior of the search space, rendering it more adaptable for exploration across diverse areas, provided a sufficient number of restarts. Nonetheless, both approaches may exhibit distinct behaviors contingent upon the characteristics of the search space. For instance, in scenarios like the barren plateau problem, where the gradient of the cost function is unreliable, the SHC-RR heuristic might prove more efficacious, while local search methods, including our heuristic model, might encounter challenges in navigating optimal solutions within intricate and expansive search spaces due to inherent assumptions about the structure of the search space \cite{R18}.

\subsection{SHC-RR}

Stochastic Hill Climbing (SHC) with random restarts employs a strategy wherein a predefined number of random restart points are introduced during execution, and each of these restart points undergoes optimization via SHC. A crucial facet of this strategy lies in conducting random restarts across all dimensions available. Put differently, each restart involves perturbing the candidate point across all dimensions of the search space randomly. Conversely, in the SHC algorithm, optimization candidates iteratively modify one dimension, chosen at random, at each step.

\begin{figure}[ht]
\centering
\includegraphics[width=6.5cm, height=4.5cm]{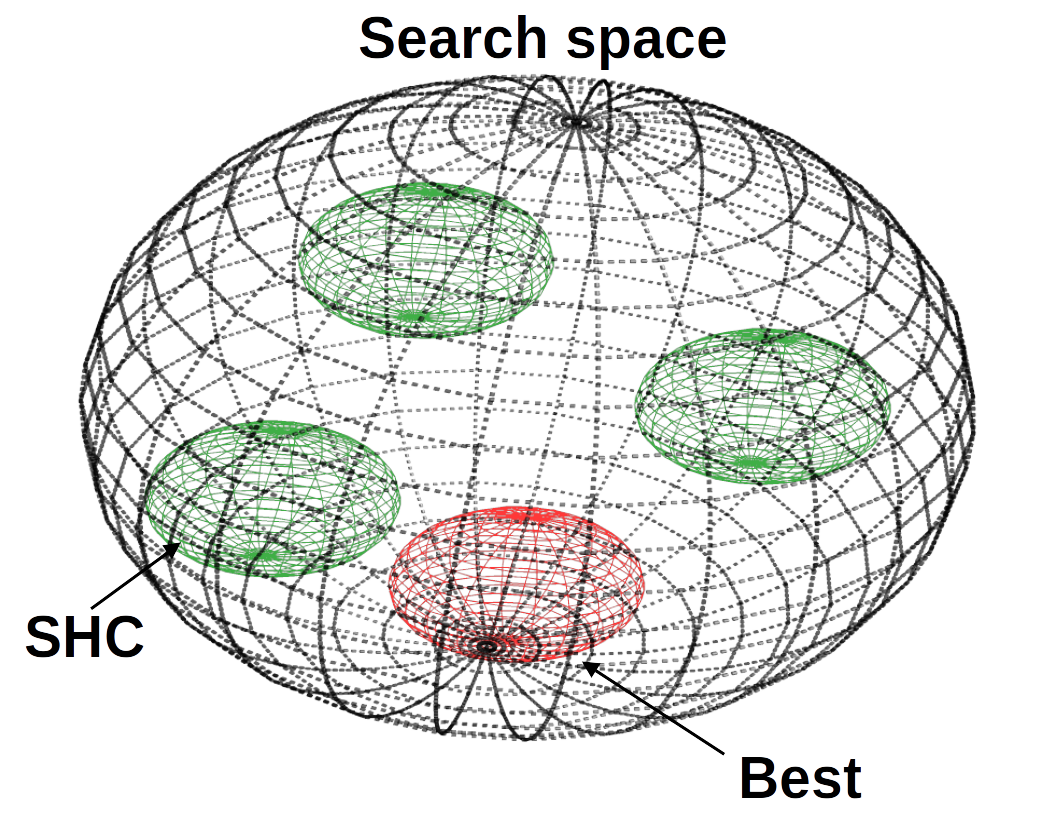}
\caption{SHC-RR visualization of exploration in a search space.}
\label{fig:shc_rr_search_space}
\end{figure}

A visualization depicting how this heuristic navigates a given search space to uncover an optimal solution is presented in \textbf{Figure \ref{fig:shc_rr_search_space}}. Here, the search space is depicted as an empty white sphere, with each green sphere representing one of the random restarts. Both the initial starting points and the restarting points throughout the heuristic's execution operate similarly. With our figure as a reference, a starting point can be any location within the sphere, while a restarting point may occupy a different position in the sphere, featuring distinct values for $x$, $y$, and $z$.

\begin{figure}[ht]
\centering
\includegraphics[width=6.5cm, height=4.5cm]{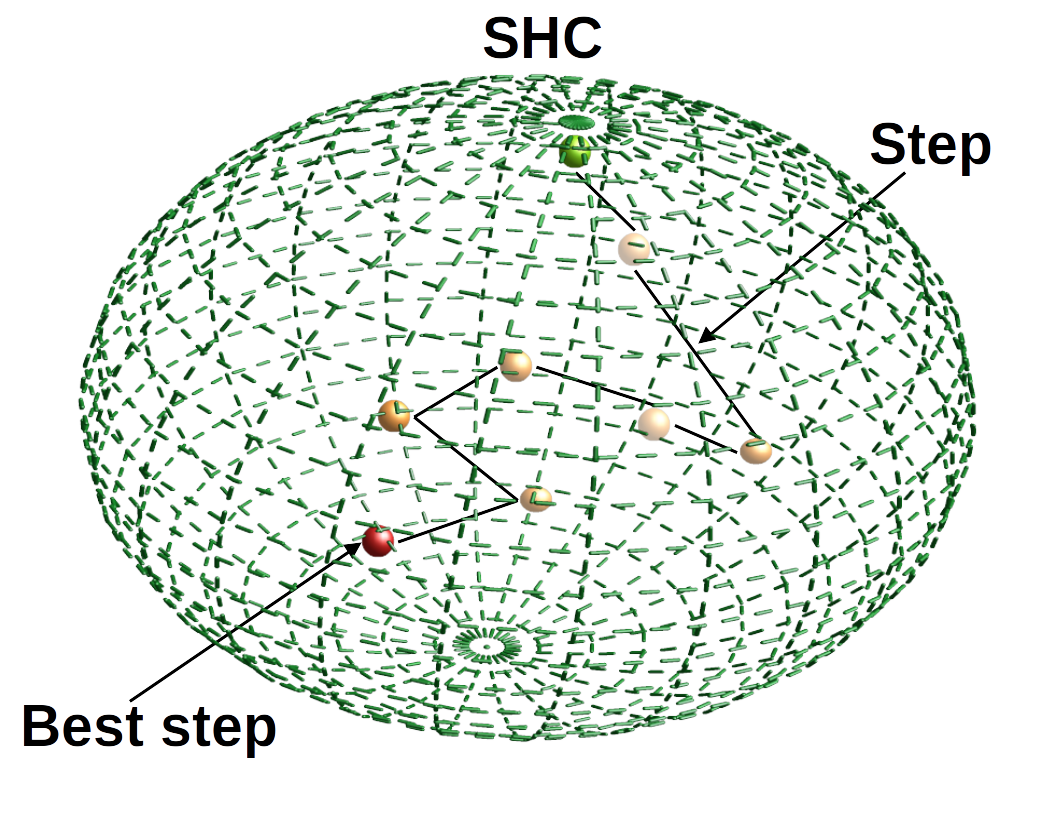}
\caption{Stochastic Hill Climbing}
\label{fig:shc_steps}
\end{figure}

Once a candidate point is situated within a green sphere (\textbf{Figure \ref{fig:shc_steps}}), the Stochastic Hill Climbing (SHC) algorithm commences, which entails making small one-dimensional (random) jumps. Specifically, we take a point characterized by certain $x$, $y$, and $z$ values, modifying only one dimension at a time for each step in SHC (with the number of steps predetermined). Upon completion of the SHC process, the best candidate is retained, and the procedure repeats a predetermined number of times (i.e., the number of restarts), with each subsequent best candidate compared to its predecessor. Ultimately, the overall best candidate of the heuristic is chosen after the process.

\begin{figure}[ht]
\centering
\includegraphics[width=15cm, height=13cm]{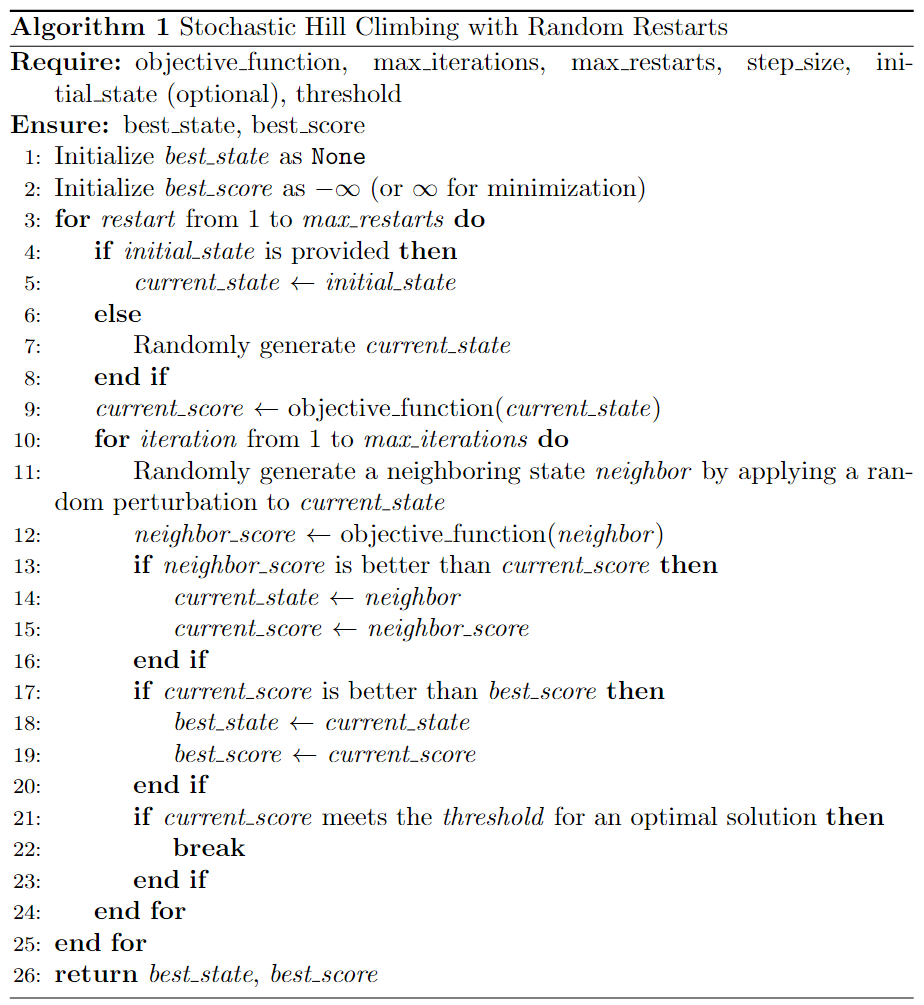}
\end{figure}

\subsection{Local search}

The Local Search heuristic closely resembles SHC-RR, differing primarily in the incorporation of a localized exploration strategy. In this instance, we define a local area or space to concentrate the exploration efforts. This local area is delineated by a perturbation parameter, facilitating jumps within the search space that are relatively proximate to each other.

\begin{figure}[ht]
\centering
\includegraphics[width=6.5cm, height=4.5cm]{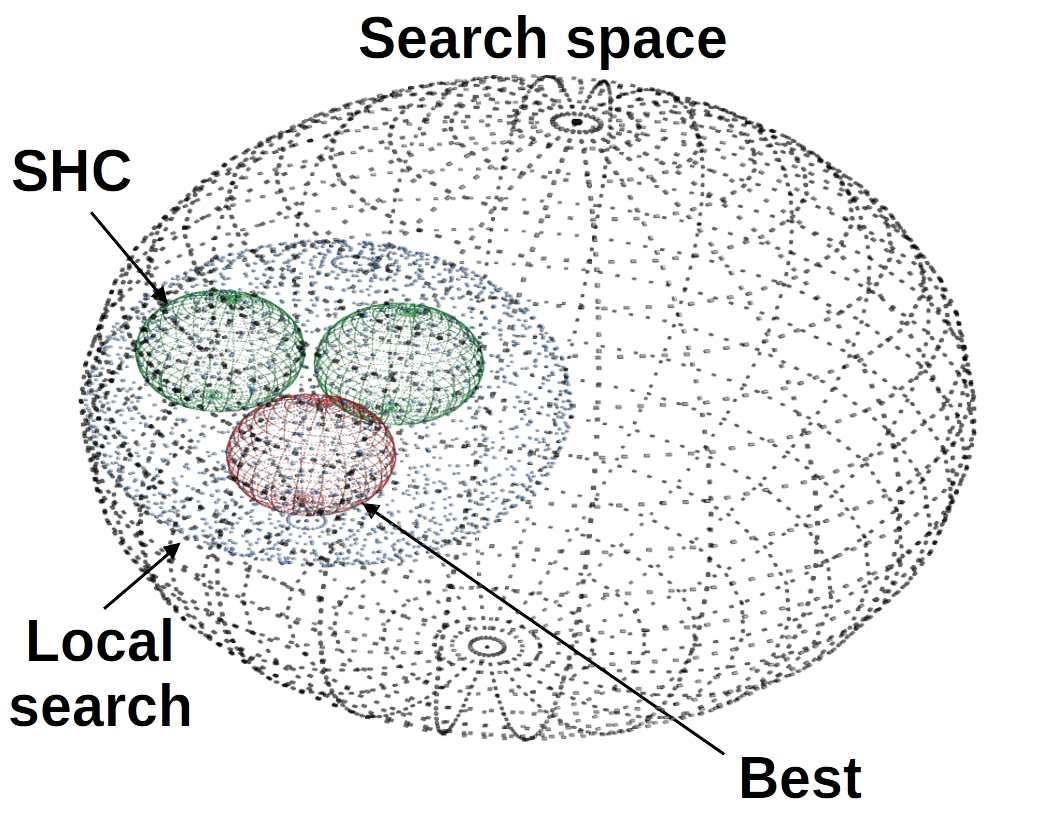}
\caption{LS visualization of exploration in a search space.}
\label{fig:local_search_search_space}
\end{figure}

Utilizing the depiction in \textbf{Figure \ref{fig:local_search_search_space}}, we can illustrate how this heuristic traverses a search space. The local space is initially positioned at a random starting point within the search space. Subsequently, the local search unfolds through a specified number of iterations of SHC, akin to the previous method, advancing the exploration utilizing the perturbation parameter of the local space. Notably, the local space isn't static; instead, it expands or shifts with each perturbation. Each perturbation propels the local search forward one dimension at a time.

\begin{figure}[ht!]
\centering
\includegraphics[width=15cm, height=18cm]{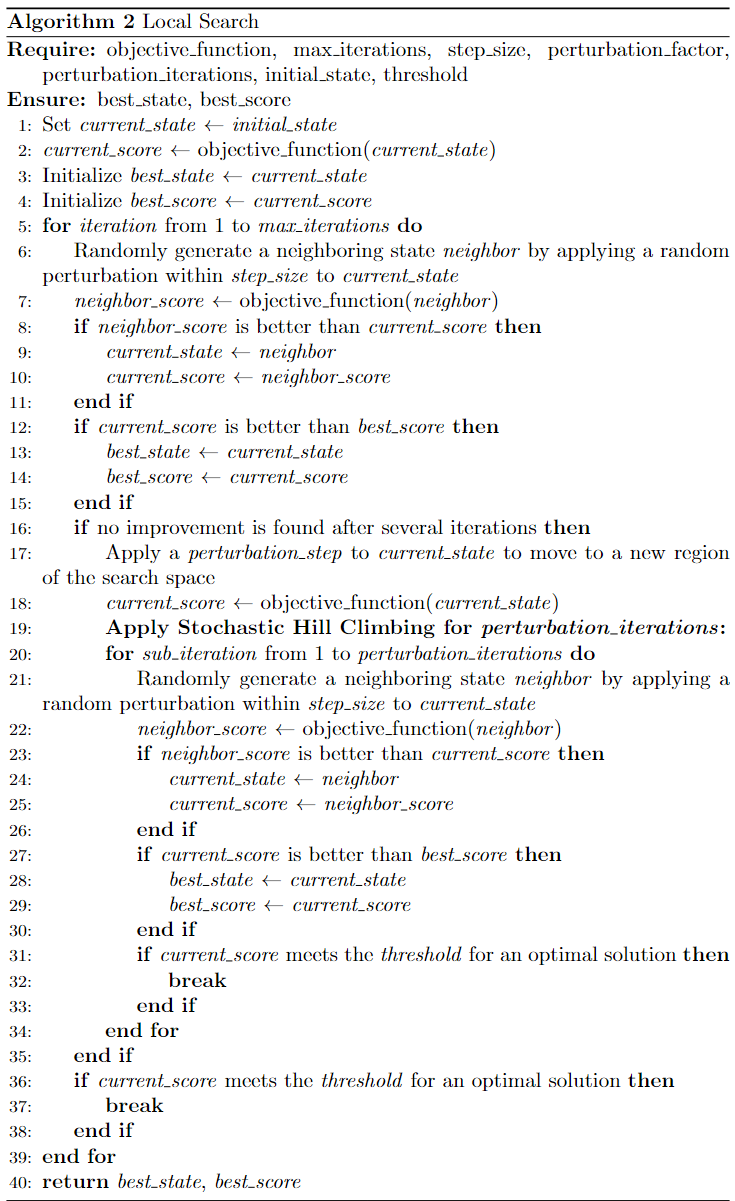}
\end{figure}

\section{Experiments and Results}

We benchmarked three QAOA depths—$1L$, $2L$, and $3L$—on two problem families: Max-Cut and the random-field Ising model (RFIM).

\paragraph{Benchmark instances}
\begin{itemize}
  \item \textbf{Max-Cut:} cyclic and complete graphs with $4$, $10$, and $15$ vertices.
  \item \textbf{RFIM:} 100 random instances.  Each node carries a local magnetic field and has at least one edge to ensure non-trivial interactions.
\end{itemize}

\paragraph{QAOA circuits}
Every layer applies a phase unitary followed by a mixing unitary formed by separate $RX$ and $RY$ rotations.  

For each depth we also evaluated a \textit{+Ent} variant that inserts an entangling block between the two single-qubit rotations.

\paragraph{Classical optimization}
Each Max-Cut circuit was optimized 100 times:  
\begin{itemize}
  \item 100 random restarts of stochastic hill climbing with random restarts (SHC-RR);
  \item 100 perturbation steps of local search, using either multiplicative (\textsc{LS}) or additive (\textsc{LS*}) updates drawn from $\mathcal{U}(-0.2,\,0.2)$;
  \item 50 simulated-annealing steps for SHC.
\end{itemize}

\paragraph{Evaluation}
For every RFIM instance we computed the expected energy value (EEV) of each circuit and reported its mean ratio to the ideal ground-state energy.

\subsection{Max-Cut problems}

The preliminary stage of our comparative analysis between Stochastic Hill Climbing with Random Restarts (SHC-RR) and Local Search (utilizing both multiplication and summation operations) commenced with an examination of max-cut problems. The tables results can are in the Appedix A section.

Across all six Max-Cut benchmarks—cyclic and complete graphs with 4, 10, and 15 nodes—\textbf{stochastic hill-climbing with random restarts (SHC-RR)} consistently achieved the lowest mean expected energy values (EEVs).  Additive local search (LS* with summation) ranked second and systematically narrowed the gap to SHC-RR, while multiplicative local search (LS) showed both higher energies and larger variances.

Circuit depth and entanglement affected performance unevenly.  The shallow, entangled $3p_{\text{ent}}$ mixer excelled only on the smallest complete graph, but underperformed elsewhere.  From six parameters upward, \emph{adding an entangling block} (\textit{ent}) either improved or matched the plain mixer, with $6p_{\text{ent}}$ dominating all larger instances: it reached the global optimum on the 4-node cyclic graph, yielded the best mean EEVs on the 10- and 15-node cyclic graphs, and produced the single best result ($\mathrm{EEV}=55.98$) on the 15-node complete graph.

Combining both trends, the empirical hierarchy is:

\[
\boxed{\text{SHC-RR} \;>\; \text{LS*} \;>\; \text{LS}}, \qquad
\boxed{6p_{\text{ent}}\text{ (or deeper)} \;>\; 6p \;>\; 3p \;>\; 3p_{\text{ent}}}.
\]

Hence, when resources permit, SHC-RR coupled with a moderately deep entangled mixer offers the most reliable route to optimal—or near-optimal—Max-Cut solutions, while LS* provides a lightweight alternative that retains most of the performance advantage.

\begin{figure}[ht]
\centering
\includegraphics[width=\textwidth]{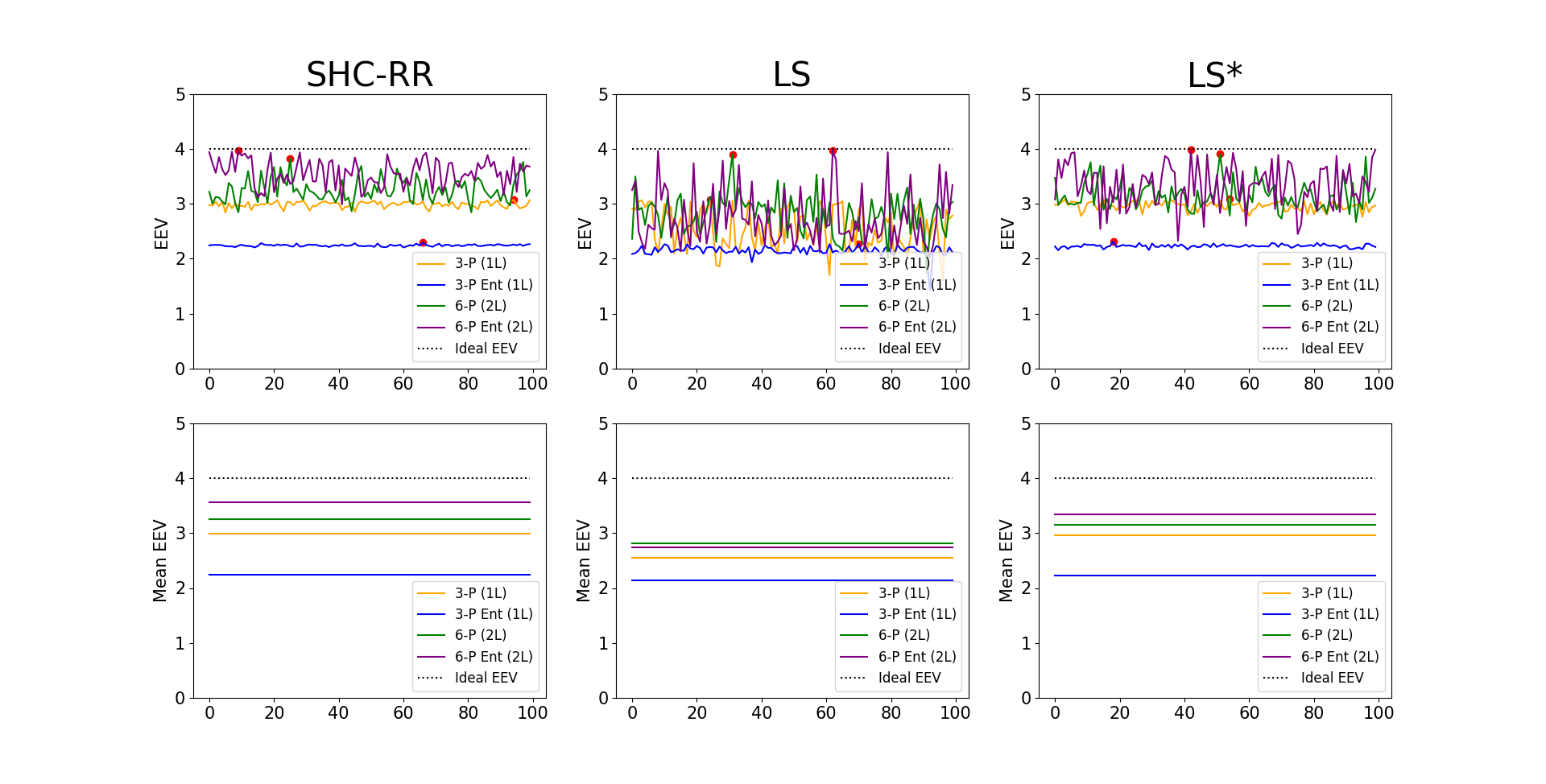}
\caption{EEV results and mean EEVs comparison for QAOA using SCH-RR, LS, and LS* (sum) in 100 experiments for the max-cut problem with 4 nodes in cyclic configuration.}
\label{fig:COMP_EEV_QAOA_4n_100_CYC}
\end{figure}

\begin{figure}[ht!]
\centering
\includegraphics[width=\textwidth]{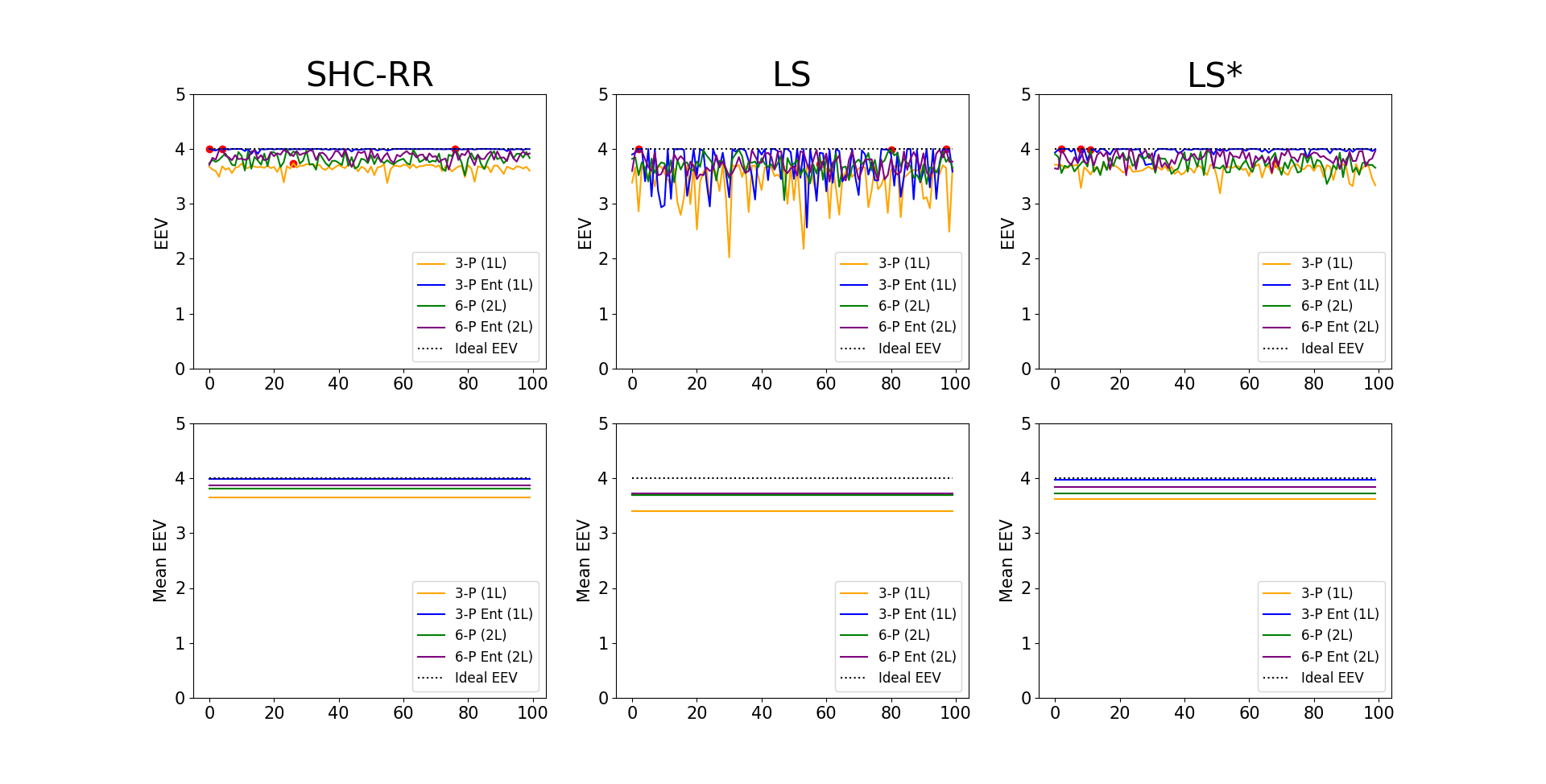}
\caption{EEV results and mean EEVs comparison for QAOA using SCH-RR, LS, and LS* (sum) in 100 experiments for the max-cut problem with 4 nodes in complete configuration.}
\label{fig:COMP_EEV_QAOA_4n_100_COM}
\end{figure}

\begin{figure}[ht!]
\centering
\includegraphics[width=\textwidth]{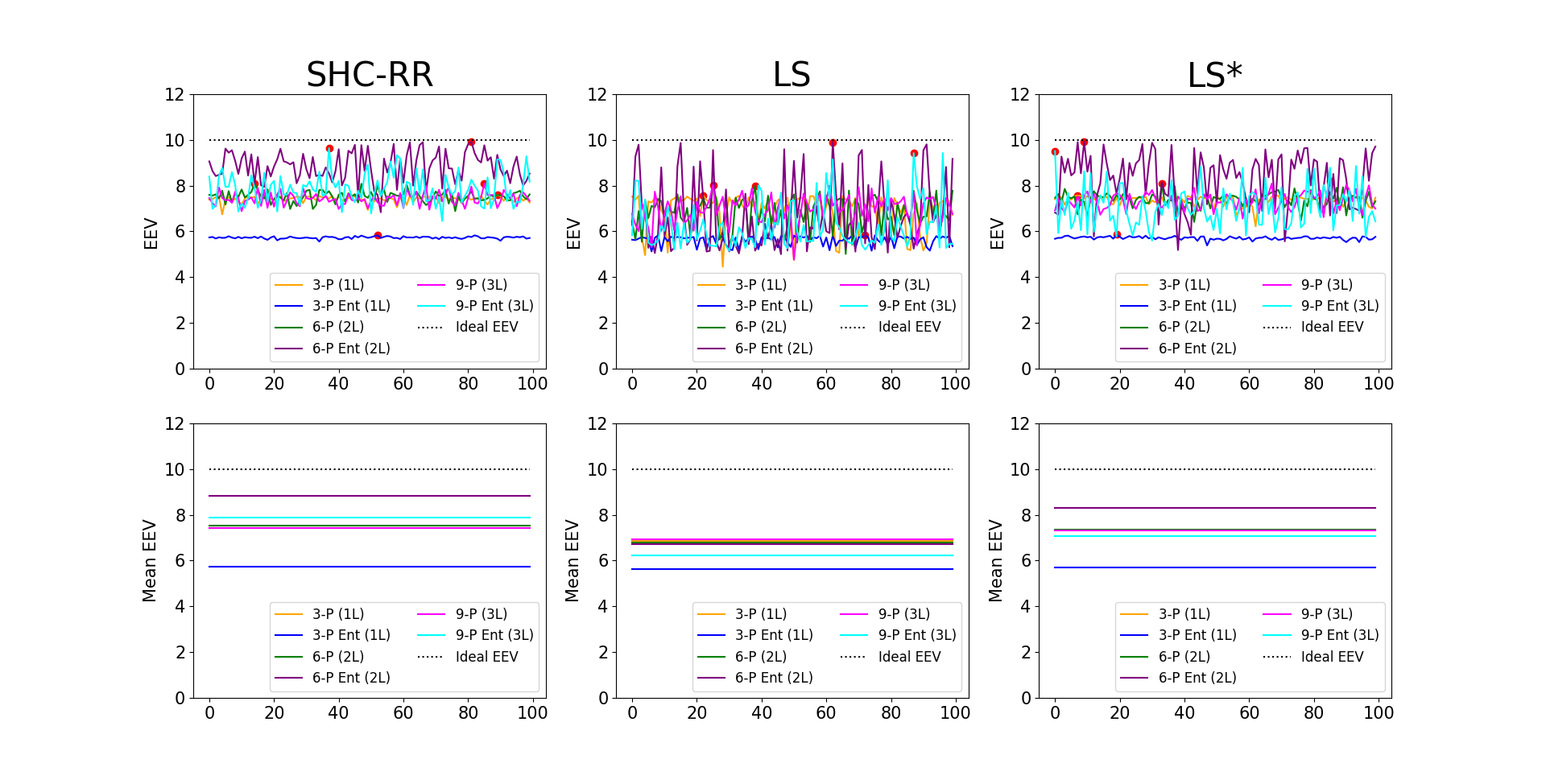}
\caption{EEV results and mean EEVs comparison for QAOA using SCH-RR, LS, and LS* (sum) in 100 experiments for the max-cut problem with 10 nodes in cyclic configuration.}
\label{fig:COMP_EEV_QAOA_10n_100_CYC}
\end{figure}

\begin{figure}[ht]
\centering
\includegraphics[width=\textwidth]{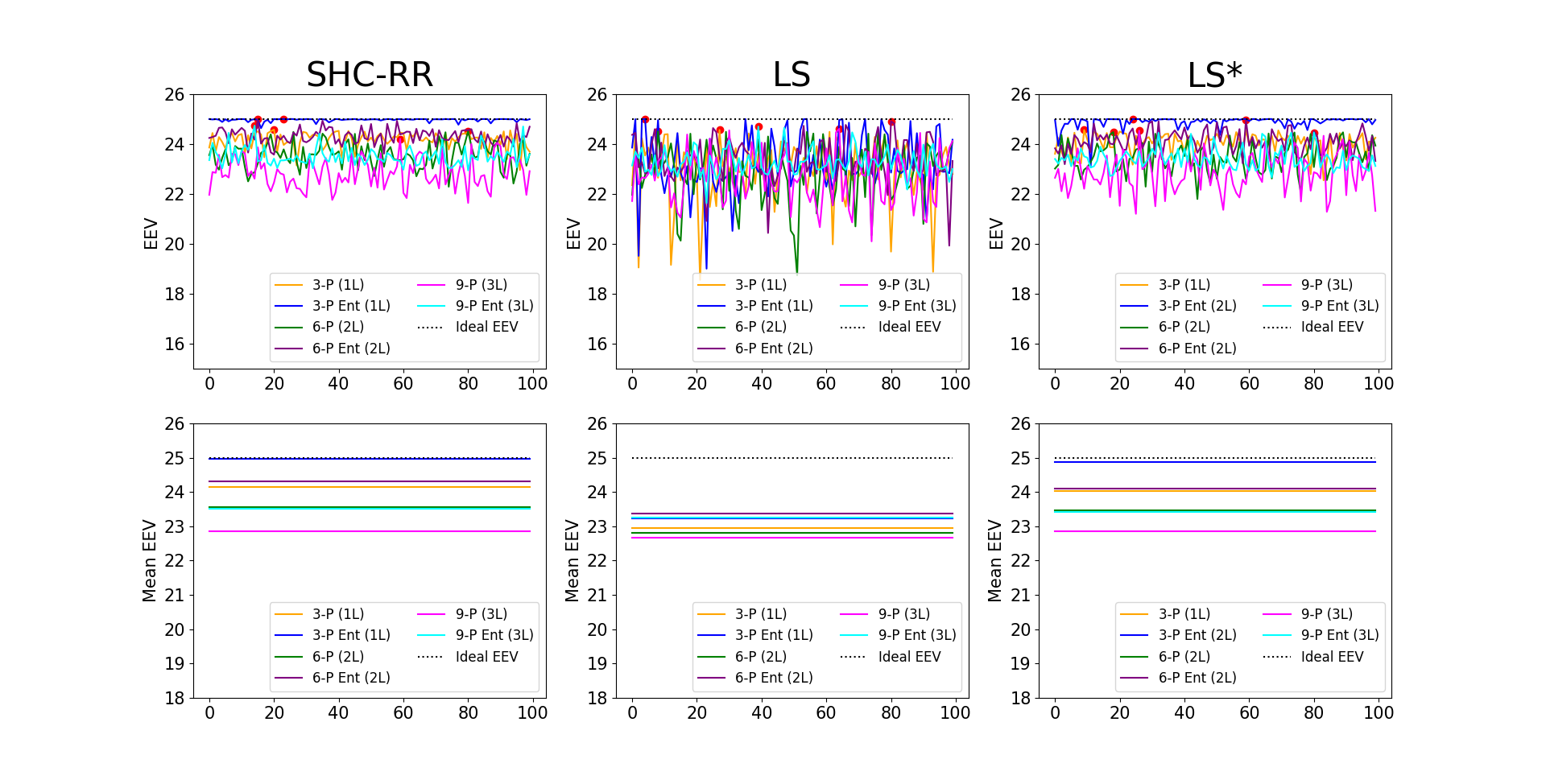}
\caption{EEV results and mean EEVs comparison for QAOA using SCH-RR, LS and LS* in 100 experiments for the max-cut problem with 10 nodes in complete configuration.}
\label{fig:COMP_EEV_QAOA_10n_100_COM}
\end{figure}

\begin{figure}[ht!]
\centering
\includegraphics[width=\textwidth]{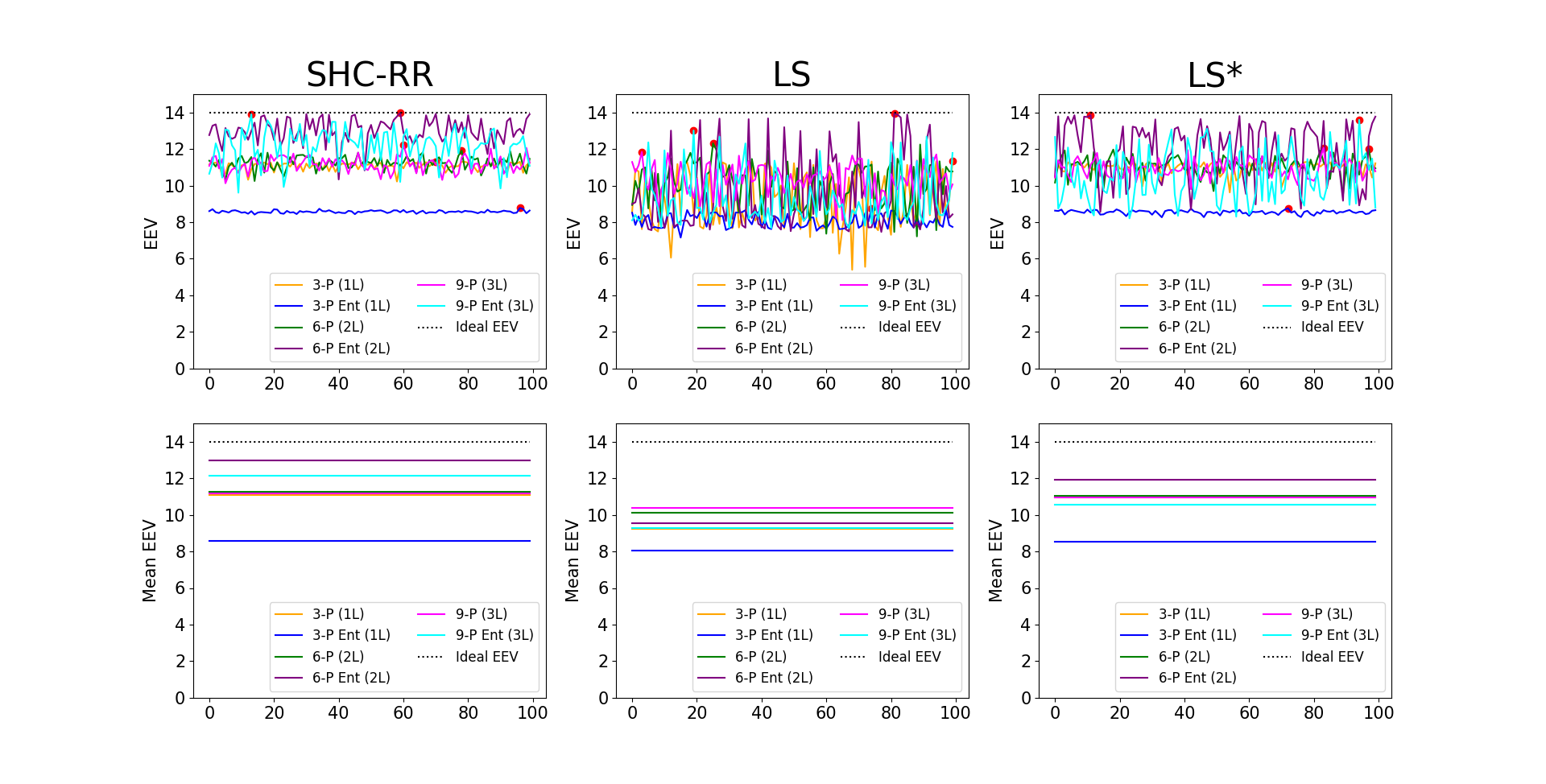}
\caption{EEV results and mean EEVs comparison for QAOA using SCH-RR and LS in 100 experiments for the max-cut problem with 15 nodes in cyclic configuration.}
\label{fig:COMP_EEV_QAOA_15n_100_CYC}
\end{figure}

\begin{figure}[ht]
\centering
\includegraphics[width=\textwidth]{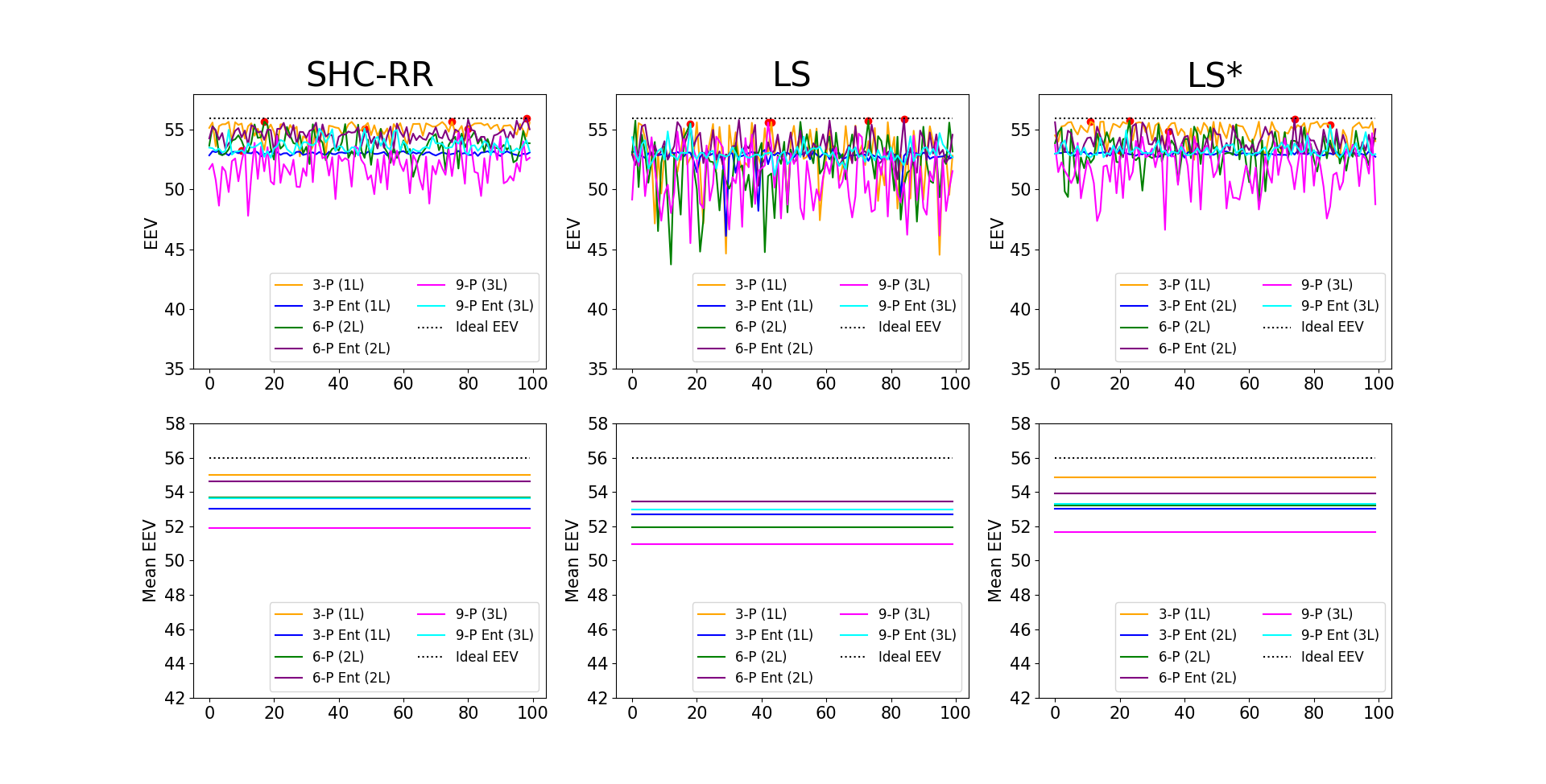}
\caption{EEV results and mean EEVs comparison for QAOA using SCH-RR and LS in 100 experiments for the max-cut problem with 15 nodes in complete configuration.}
\label{fig:COMP_EEV_QAOA_15n_100_COM}
\end{figure}

\subsection{Random ising model problems}

In the second part of our analysis, we delve into the results of the random Ising model problems solved using the SHC-RR approach with various QAOA models: $1L$ with $3$ parameters, $2L$ with $6$ parameters, and $3L$ with $9$ parameters. Each model was tested both with and without the presence of entanglement stages.

Over 100 random-field Ising instances, \textbf{SHC-RR} again produced the energies closest to the exact ground state, recording a mean deviation of $-0.97$ for the $3p$ mixer (Table~\ref{tab:EEVs_diff_100_exp}).  LS* ranked second ($-1.35$), while LS showed the largest gaps and variances; the poorest case was LS with the $9p_{\text{ent}}$ circuit ($-4.98$).

Performance deteriorated systematically with depth: the \emph{shallow} $3p$ mixers (1~layer)—entangled or not—were consistently the top performers, whereas both $6p$ and $9p$ circuits yielded increasingly negative deviations.  Entangling the mixer was generally detrimental; the lone exception was the $3p_{\text{ent}}$ circuit, whose accuracy matched its non-entangled counterpart.

Local-search trajectories exhibited pronounced spikes, isolated runs that approached the ideal energy, highlighting a strong dependence on the initial point. SHC-RR, by contrast, delivered both the best means and the tightest distributions, confirming the optimizer hierarchy:

\[
\boxed{\text{SHC-RR} \;>\; \text{LS*} \;>\; \text{LS}}
\quad\text{and}\quad
\boxed{3p\;(\text{or}\;3p_{\text{ent}})\;>\;6p\;>\;9p}.
\]

\begin{figure}[ht!]
\centering
\includegraphics[width=\textwidth]{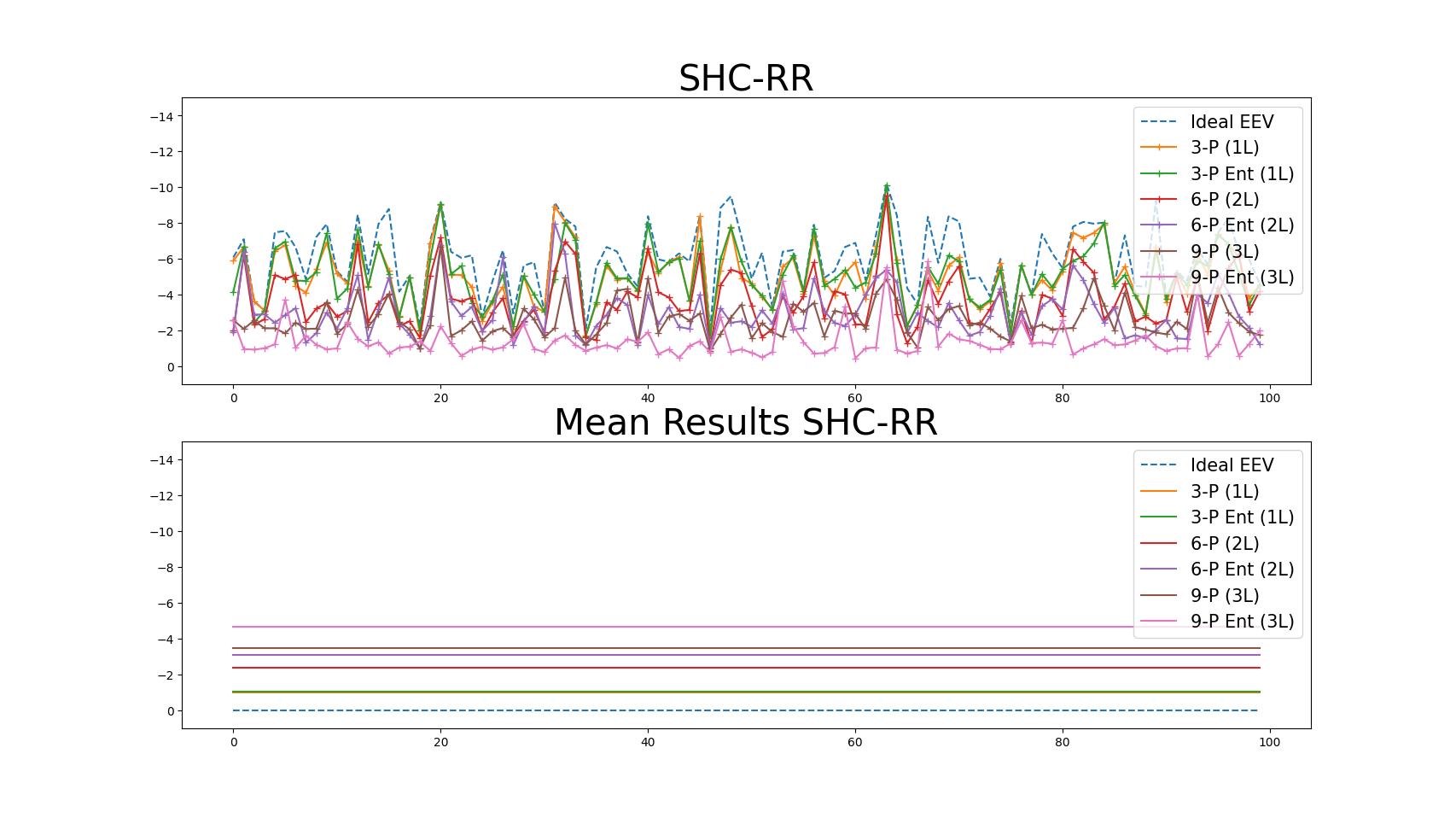}
\caption{EEV results and mean EEVs for the random Ising model problems using SHC-RR with 100 restarts, implementing different QAOA models.}
\label{fig:Rand_Ising_SHC-RR_100rest}
\end{figure}

\begin{figure}[ht!]
\centering
\includegraphics[width=\textwidth]{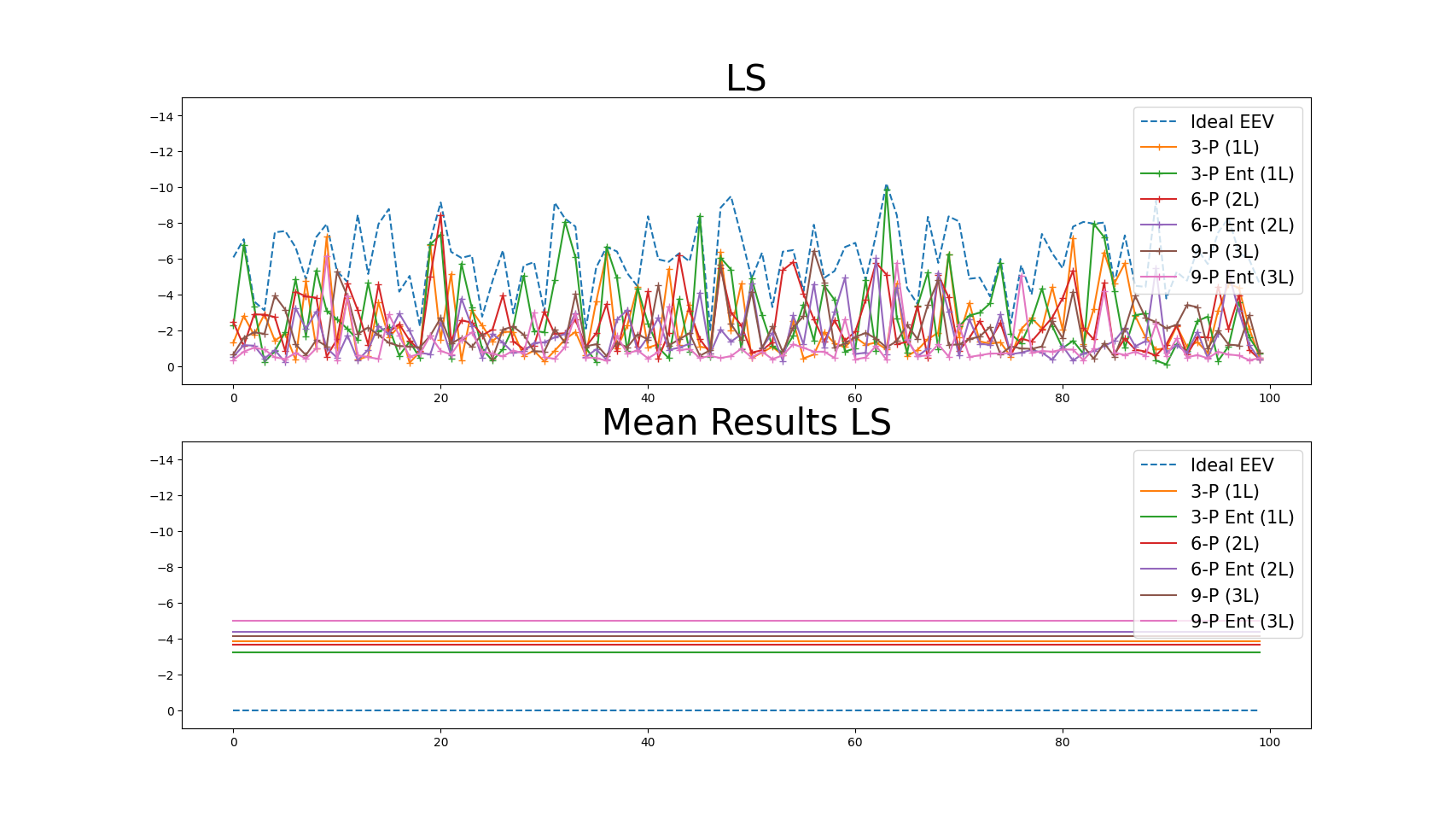}
\caption{EEV results and mean EEVs for the random Ising model problems using LS with 100 restarts, implementing different QAOA models.}
\label{fig:Rand_Ising_LS_100rest}
\end{figure}

\begin{figure}[ht!]
\centering
\includegraphics[width=\textwidth]{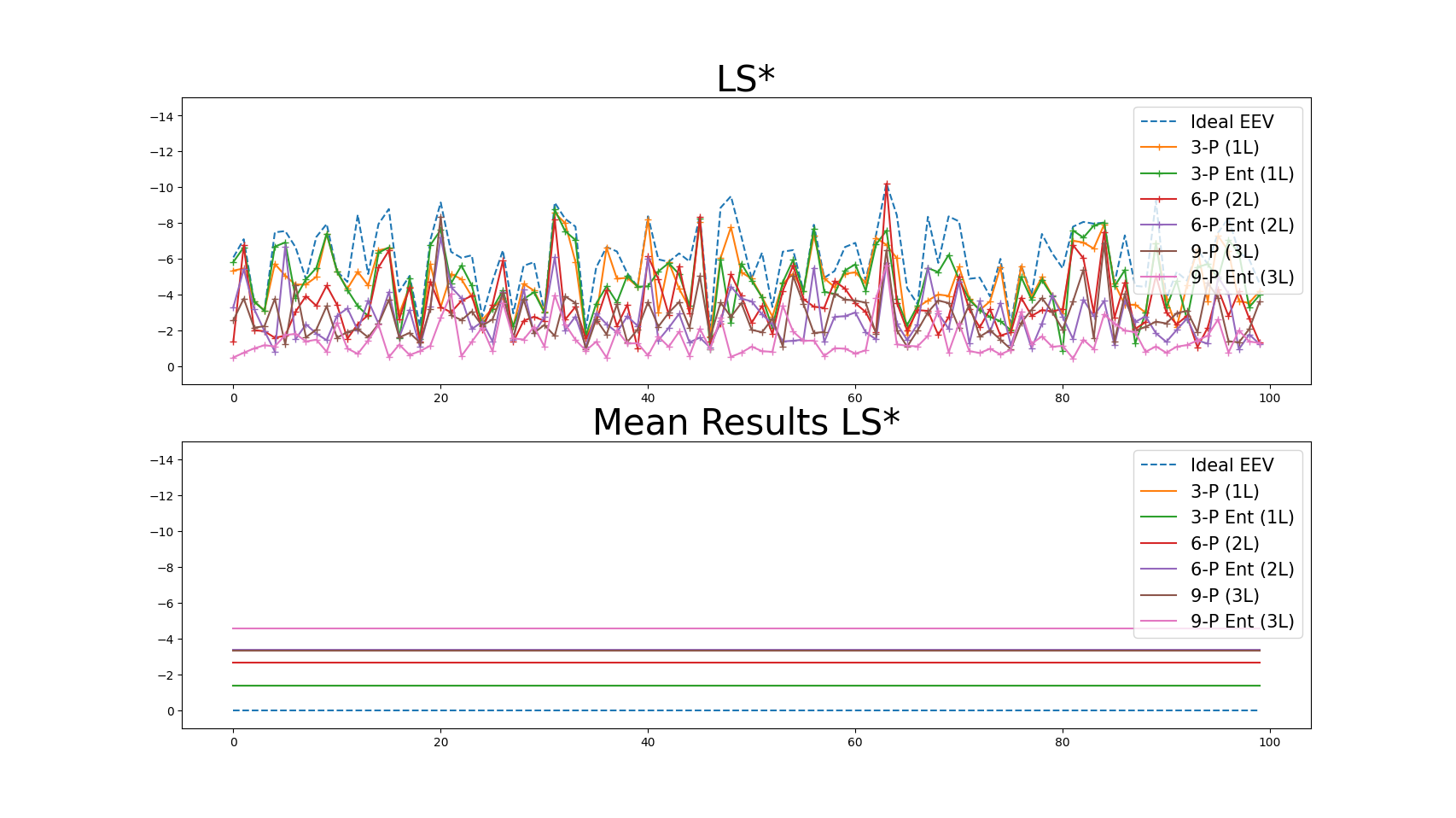}
\caption{EEV results and mean EEVs for the random Ising model problems using LS* (using summation) with 100 restarts, implementing different QAOA models.}
\label{fig:Rand_Ising_LS_Sum_100rest}
\end{figure}

\begin{table}[ht!]
\centering
\caption{EEVs mean difference with respect to the ideal EEVs for the SHC-RR, LS, and LS* approaches, using 100 restarts (or perturbations) iterations.}
\label{tab:EEVs_diff_100_exp}
\begin{tabular}{|c|c|c|c|}
\hline
       & SHC-RR             & LS                 & LS*                \\ \hline
Model  & Mean diff          & Mean diff          & Mean diff          \\ \hline
3p     & \textbf{-0.966828} & -3.827414          & \textbf{-1.347080} \\ \hline
3p ent & -1.031253          & \textbf{-3.210375} & -1.374083          \\ \hline
6p     & -2.365603          & -3.634124          & -2.631068          \\ \hline
6p ent & -3.088031          & -4.370483          & -3.353920          \\ \hline
9p     & -3.460539          & -4.148047          & -3.299458          \\ \hline
9p ent & -4.652033          & -4.975057          & -4.576437          \\ \hline
\end{tabular}
\end{table}

\section{Conclusions}

This study has demonstrated that the Stochastic Hill Climbing with Random Restarts (SHC-RR) approach consistently outperforms both the Local Search (LS with multiplication) and LS* (LS with summation) in achieving superior Expected Energy Values (EEVs) across a range of analyzed problems. This superiority persists despite all strategies undergoing an identical number of iterations, underscoring SHC-RR's robustness in various test scenarios.

The efficacy of the LS approach is notably influenced by the choice of operation (multiplication or summation) employed in the perturbation step, which significantly impacts the strategy's overall performance.

Operational efficiency remains consistent across the tested strategies, thanks to uniform iteration counts for restarts or perturbations. This consistency is crucial, especially as the number of parameters increases, ensuring comparable execution times across different QAOA depths.

In terms of model performance, the $2L$ entangled model (with six parameters) excels in cyclic configuration max-cut problems, whereas the $1L$ entangled model (with three parameters) shows superior efficacy in complete configuration max-cut problems. Interestingly, for max-cut problems, entangled models generally yield better average outcomes, hinting at the effectiveness of maintaining low-depth QAOA models for quality results.

In random Ising model problems, the $1L$ models outperform others across all approaches, highlighting the significant impact of search space dimensionality on performance, particularly under fixed iteration counts.

Contrary to findings in max-cut problems, entangled models underperformed relative to their non-entangled counterparts in random Ising model scenarios. This discrepancy underscores the nuanced impact of entanglement on optimization outcomes, warranting further investigation.

Remarkably, SHC-RR's success, despite its relatively simple design devoid of a tailored exploration strategy, underscores the effectiveness of a more exploratory approach over predetermined search strategies. This observation suggests that an open exploration of the search space, without preconceived biases, is advantageous for the generation of optimal parameters and EEV outcomes.

\section{Future work}

For future work, exploring additional heuristics could provide valuable insights into optimization performance.

Additionally, exploring the impact of varying parameters, such as the number of iterations or the size of the search space, could provide deeper insights into the behavior and performance of LS heuristics. Understanding how these factors influence optimization outcomes can inform the development of more effective and efficient optimization strategies.

\section*{Acknowledgements}
We acknowledge the partial support of projects 20240220 and 20240421-SIP-IPN, Mexico. S.H. Dong started this work on the leave of IPN due to permission for a research stay in China.
\bibliographystyle{IEEEtran}
\bibliography{references}

\appendix
\section{Appendix}
\subsection{Max-cut tables results}

In this section of the Appedix, we present the tables for the max-cut experiments in the different QAOA models and optimization strategies.

\begin{table}[ht!]
\centering
\caption{EEV best, mean and variance for the 4-node cyclic max-cut problem solved using SHC-RR, LS, and LS* (sum) with different QAOA models, $3p$ and $3p$ entangled corresponds to the $1L$ of depth model, and $6p$ and $6p$ entangled corresponds to the $2L$ of depth model.}
\label{tab:comp_shc-rr_vs_ls_4n_cyc}
\begin{tabular}{|c|ccc|ccc|ccc|}
\hline
       & \multicolumn{3}{c|}{SHC-RR}                                                          & \multicolumn{3}{c|}{LS}                                                              & \multicolumn{3}{c|}{LS*}                                                             \\ \hline
Model  & \multicolumn{1}{c|}{Best}            & \multicolumn{1}{c|}{Mean}            & Var    & \multicolumn{1}{c|}{Best}            & \multicolumn{1}{c|}{Mean}            & Var    & \multicolumn{1}{c|}{Best}            & \multicolumn{1}{c|}{Mean}            & Var    \\ \hline
3p     & \multicolumn{1}{c|}{3.0742}          & \multicolumn{1}{c|}{2.9882}          & 0.0028 & \multicolumn{1}{c|}{3.0742}          & \multicolumn{1}{c|}{2.5550}          & 0.1403 & \multicolumn{1}{c|}{3.0917}          & \multicolumn{1}{c|}{2.9625}          & 0.0062 \\ \hline
3p ent & \multicolumn{1}{c|}{2.2949}          & \multicolumn{1}{c|}{2.2426}          & 0.0003 & \multicolumn{1}{c|}{2.2753}          & \multicolumn{1}{c|}{2.1461}          & 0.0089 & \multicolumn{1}{c|}{2.3105}          & \multicolumn{1}{c|}{2.2317}          & 0.0008 \\ \hline
6p     & \multicolumn{1}{c|}{3.8320}          & \multicolumn{1}{c|}{3.2564}          & 0.0438 & \multicolumn{1}{c|}{3.9042}          & \multicolumn{1}{c|}{\textbf{2.8074}} & 0.1262 & \multicolumn{1}{c|}{3.9101}          & \multicolumn{1}{c|}{3.1539}          & 0.0620 \\ \hline
6p ent & \multicolumn{1}{c|}{\textbf{3.9726}} & \multicolumn{1}{c|}{\textbf{3.5658}} & 0.0566 & \multicolumn{1}{c|}{\textbf{3.9746}} & \multicolumn{1}{c|}{2.7459}          & 0.2357 & \multicolumn{1}{c|}{\textbf{3.9882}} & \multicolumn{1}{c|}{\textbf{3.3484}} & 0.1499 \\ \hline
\end{tabular}
\end{table}

\begin{table}[ht!]
\centering
\caption{EEV best, mean and variance for the 4-node complete max-cut problem solved using SHC-RR, LS, and LS* (sum) with different QAOA models, $3p$ and $3p$ entangled corresponds to the $1L$ of depth model, and $6p$ and $6p$ entangled corresponds to the $2L$ of depth model.}
\label{tab:comp_shc-rr_vs_ls_4n_com}
\begin{tabular}{|c|ccc|ccc|ccc|}
\hline
       & \multicolumn{3}{c|}{SHC-RR}                                                       & \multicolumn{3}{c|}{LS}                                                           & \multicolumn{3}{c|}{LS*}                                                          \\ \hline
Model  & \multicolumn{1}{c|}{Best}         & \multicolumn{1}{c|}{Mean}            & Var    & \multicolumn{1}{c|}{Best}         & \multicolumn{1}{c|}{Mean}            & Var    & \multicolumn{1}{c|}{Best}         & \multicolumn{1}{c|}{Mean}            & Var    \\ \hline
3p     & \multicolumn{1}{c|}{3.7412}       & \multicolumn{1}{c|}{3.6536}          & 0.0045 & \multicolumn{1}{c|}{3.7226}       & \multicolumn{1}{c|}{3.4052}          & 0.1321 & \multicolumn{1}{c|}{3.7431}       & \multicolumn{1}{c|}{3.6241}          & 0.0101 \\ \hline
3p ent & \multicolumn{1}{c|}{\textbf{4.0}} & \multicolumn{1}{c|}{\textbf{3.9927}} & 0.0002 & \multicolumn{1}{c|}{\textbf{4.0}} & \multicolumn{1}{c|}{3.6937}          & 0.1276 & \multicolumn{1}{c|}{\textbf{4.0}} & \multicolumn{1}{c|}{\textbf{3.9687}} & 0.0039 \\ \hline
6p     & \multicolumn{1}{c|}{3.9970}       & \multicolumn{1}{c|}{3.8063}          & 0.0103 & \multicolumn{1}{c|}{3.9921}       & \multicolumn{1}{c|}{3.6959}          & 0.0272 & \multicolumn{1}{c|}{3.9804}       & \multicolumn{1}{c|}{3.7297}          & 0.0169 \\ \hline
6p ent & \multicolumn{1}{c|}{3.9960}       & \multicolumn{1}{c|}{3.8720}          & 0.0058 & \multicolumn{1}{c|}{3.9941}       & \multicolumn{1}{c|}{\textbf{3.7208}} & 0.0275 & \multicolumn{1}{c|}{\textbf{4.0}} & \multicolumn{1}{c|}{3.8334}          & 0.0126 \\ \hline
\end{tabular}
\end{table}

\begin{table}[ht!]
\centering
\caption{EEV best, mean and variance for the 10-node cyclic max-cut problem solved using SHC-RR, LS, and LS* (sum) with different QAOA models, $3p$ and $3p$ entangled corresponds to the $1L$ of depth model, $6p$ and $6p$ entangled corresponds to the $2L$ of depth model, $9p$ and $9p$ entangled corresponds to the $3L$ of depth model.}
\label{tab:comp_shc-rr_vs_ls_10n_cyc}
\begin{tabular}{|c|ccc|ccc|cc|c|}
\hline
       & \multicolumn{3}{c|}{SHC-RR}                                                          & \multicolumn{3}{c|}{LS}                                                              & \multicolumn{2}{c|}{LS*}                               &        \\ \hline
Model  & \multicolumn{1}{c|}{Best}            & \multicolumn{1}{c|}{Mean}            & Var    & \multicolumn{1}{c|}{Best}            & \multicolumn{1}{c|}{Mean}            & Var    & \multicolumn{1}{c|}{Best}            & Mean            & Var    \\ \hline
3p     & \multicolumn{1}{c|}{7.5917}          & \multicolumn{1}{c|}{7.4103}          & 0.0176 & \multicolumn{1}{c|}{7.5644}          & \multicolumn{1}{c|}{6.8746}          & 0.6118 & \multicolumn{1}{c|}{7.5644}          & 7.3558          & 0.0435 \\ \hline
3p ent & \multicolumn{1}{c|}{5.8222}          & \multicolumn{1}{c|}{5.7242}          & 0.0022 & \multicolumn{1}{c|}{5.8359}          & \multicolumn{1}{c|}{5.6113}          & 0.0432 & \multicolumn{1}{c|}{5.8535}          & 5.7061          & 0.0047 \\ \hline
6p     & \multicolumn{1}{c|}{8.0898}          & \multicolumn{1}{c|}{7.5307}          & 0.0542 & \multicolumn{1}{c|}{7.9804}          & \multicolumn{1}{c|}{6.8013}          & 0.4878 & \multicolumn{1}{c|}{8.0957}          & 7.3691          & 0.1045 \\ \hline
6p ent & \multicolumn{1}{c|}{\textbf{9.9121}} & \multicolumn{1}{c|}{\textbf{8.8240}} & 0.4129 & \multicolumn{1}{c|}{\textbf{9.9003}} & \multicolumn{1}{c|}{6.7112}          & 2.2462 & \multicolumn{1}{c|}{\textbf{9.9160}} & \textbf{8.3009} & 1.2482 \\ \hline
9p     & \multicolumn{1}{c|}{8.0839}          & \multicolumn{1}{c|}{7.4284}          & 0.0731 & \multicolumn{1}{c|}{8.0214}          & \multicolumn{1}{c|}{\textbf{6.9448}} & 0.3583 & \multicolumn{1}{c|}{8.1777}          & 7.3274          & 0.1290 \\ \hline
9p ent & \multicolumn{1}{c|}{9.6347}          & \multicolumn{1}{c|}{7.8657}          & 0.4333 & \multicolumn{1}{c|}{9.4218}          & \multicolumn{1}{c|}{6.2282}          & 1.0277 & \multicolumn{1}{c|}{9.5097}          & 7.0739          & 0.7756 \\ \hline
\end{tabular}
\end{table}

\begin{table}[ht!]
\centering
\caption{EEV best, mean and variance for the 10-node complete max-cut problem solved using SHC-RR, LS, and LS* with different QAOA models, $3p$ and $3p$ entangled corresponds to the $1L$ of depth model, $6p$ and $6p$ entangled corresponds to the $2L$ of depth model, $9p$ and $9p$ entangled corresponds to the $3L$ of depth model.}
\label{tab:comp_shc-rr_vs_ls_10n_com}
\begin{tabular}{|c|ccc|ccc|ccc|}
\hline
       & \multicolumn{3}{c|}{SHC-RR}                                                         & \multicolumn{3}{c|}{LS}                                                             & \multicolumn{3}{c|}{LS*}                                                            \\ \hline
Model  & \multicolumn{1}{c|}{Best}          & \multicolumn{1}{c|}{Mean}             & Var    & \multicolumn{1}{c|}{Best}          & \multicolumn{1}{c|}{Mean}             & Var    & \multicolumn{1}{c|}{Best}          & \multicolumn{1}{c|}{Mean}             & Var    \\ \hline
3p     & \multicolumn{1}{c|}{24.5751}       & \multicolumn{1}{c|}{24.1520}          & 0.1081 & \multicolumn{1}{c|}{24.5615}       & \multicolumn{1}{c|}{22.9499}          & 1.5490 & \multicolumn{1}{c|}{24.5605}       & \multicolumn{1}{c|}{24.0172}          & 0.1757 \\ \hline
3p ent & \multicolumn{1}{c|}{\textbf{25.0}} & \multicolumn{1}{c|}{\textbf{24.9675}} & 0.0034 & \multicolumn{1}{c|}{\textbf{25.0}} & \multicolumn{1}{c|}{23.2322}          & 1.3912 & \multicolumn{1}{c|}{\textbf{25.0}} & \multicolumn{1}{c|}{\textbf{24.8797}} & 0.0620 \\ \hline
6p     & \multicolumn{1}{c|}{24.5195}       & \multicolumn{1}{c|}{23.5543}          & 0.2466 & \multicolumn{1}{c|}{24.5058}       & \multicolumn{1}{c|}{22.8156}          & 1.3612 & \multicolumn{1}{c|}{24.4746}       & \multicolumn{1}{c|}{23.4665}          & 0.3540 \\ \hline
6p ent & \multicolumn{1}{c|}{24.9873}       & \multicolumn{1}{c|}{24.3117}          & 0.088  & \multicolumn{1}{c|}{24.9111}       & \multicolumn{1}{c|}{\textbf{23.3796}} & 0.7217 & \multicolumn{1}{c|}{24.9550}       & \multicolumn{1}{c|}{24.1069}          & 0.1788 \\ \hline
9p     & \multicolumn{1}{c|}{24.2041}       & \multicolumn{1}{c|}{22.8439}          & 0.3558 & \multicolumn{1}{c|}{24.6064}       & \multicolumn{1}{c|}{22.6609}          & 1.0121 & \multicolumn{1}{c|}{24.5351}       & \multicolumn{1}{c|}{22.8492}          & 0.5782 \\ \hline
9p ent & \multicolumn{1}{c|}{24.7324}       & \multicolumn{1}{c|}{23.5217}          & 0.1375 & \multicolumn{1}{c|}{24.6972}       & \multicolumn{1}{c|}{23.2611}          & 0.2533 & \multicolumn{1}{c|}{24.4599}       & \multicolumn{1}{c|}{23.4090}          & 0.1623 \\ \hline
\end{tabular}
\end{table}

\begin{table}[ht!]
\centering
\caption{EEV best, mean and variance for the 15-node cyclic max-cut problem solved using SHC-RR, LS and LS* with different QAOA models, $3p$ and $3p$ entangled corresponds to the $1L$ of depth model, $6p$ and $6p$ entangled corresponds to the $2L$ of depth model, $9p$ and $9p$ entangled corresponds to the $3L$ of depth model.}
\label{tab:comp_shc-rr_vs_ls_15n_cyc}
\begin{tabular}{|c|ccl|ccl|ccl|}
\hline
       & \multicolumn{3}{c|}{SHC-RR}                                                            & \multicolumn{3}{c|}{LS}                                                                & \multicolumn{3}{c|}{LS*}                                                               \\ \hline
Model  & \multicolumn{1}{c|}{Best}             & \multicolumn{1}{c|}{Mean}             & Var    & \multicolumn{1}{c|}{Best}             & \multicolumn{1}{c|}{Mean}             & Var    & \multicolumn{1}{c|}{Best}             & \multicolumn{1}{c|}{Mean}             & Var    \\ \hline
3p     & \multicolumn{1}{c|}{11.3105}          & \multicolumn{1}{c|}{11.0735}          & 0.0452 & \multicolumn{1}{c|}{11.3554}          & \multicolumn{1}{c|}{9.2277}           & 2.3700 & \multicolumn{1}{c|}{11.3613}          & \multicolumn{1}{c|}{10.9678}          & 0.1204 \\ \hline
3p ent & \multicolumn{1}{c|}{8.7734}           & \multicolumn{1}{c|}{8.5696}           & 0.0046 & \multicolumn{1}{c|}{8.7558}           & \multicolumn{1}{c|}{8.0617}           & 0.1297 & \multicolumn{1}{c|}{8.7578}           & \multicolumn{1}{c|}{8.5488}           & 0.0103 \\ \hline
6p     & \multicolumn{1}{c|}{11.9082}          & \multicolumn{1}{c|}{11.2449}          & 0.1234 & \multicolumn{1}{c|}{12.3085}          & \multicolumn{1}{c|}{10.1026}          & 1.4059 & \multicolumn{1}{c|}{11.9921}          & \multicolumn{1}{c|}{11.0425}          & 0.2651 \\ \hline
6p ent & \multicolumn{1}{c|}{\textbf{13.9707}} & \multicolumn{1}{c|}{\textbf{12.9881}} & 0.4279 & \multicolumn{1}{c|}{\textbf{13.9257}} & \multicolumn{1}{c|}{9.5255}           & 4.0563 & \multicolumn{1}{c|}{\textbf{13.8535}} & \multicolumn{1}{c|}{\textbf{11.9099}} & 2.2050 \\ \hline
9p     & \multicolumn{1}{c|}{12.2285}          & \multicolumn{1}{c|}{11.1598}          & 0.1621 & \multicolumn{1}{c|}{11.8398}          & \multicolumn{1}{c|}{\textbf{10.3785}} & 0.6908 & \multicolumn{1}{c|}{12.0585}          & \multicolumn{1}{c|}{10.9399}          & 0.2261 \\ \hline
9p ent & \multicolumn{1}{c|}{13.8925}          & \multicolumn{1}{c|}{12.1546}          & 0.8441 & \multicolumn{1}{c|}{13.0117}          & \multicolumn{1}{c|}{9.2915}           & 1.9725 & \multicolumn{1}{c|}{13.5957}          & \multicolumn{1}{c|}{10.5590}          & 1.9574 \\ \hline
\end{tabular}
\end{table}

\begin{table}[ht!]
\centering
\caption{EEV best, mean and variance for the 15-node complete max-cut problem solved using SHC-RR, LS and LS* with different QAOA models, $3p$ and $3p$ entangled corresponds to the $1L$ of depth model, $6p$ and $6p$ entangled corresponds to the $2L$ of depth model, $9p$ and $9p$ entangled corresponds to the $3L$ of depth model.}
\label{tab:comp_shc-rr_vs_ls_15n_com}
\begin{tabular}{|c|ccl|ccl|ccl|}
\hline
       & \multicolumn{3}{c|}{SHC-RR}                                                            & \multicolumn{3}{c|}{LS}                                                                & \multicolumn{3}{c|}{LS*}                                                               \\ \hline
Model  & \multicolumn{1}{c|}{Best}             & \multicolumn{1}{c|}{Mean}             & Var    & \multicolumn{1}{c|}{Best}             & \multicolumn{1}{c|}{Mean}             & Var    & \multicolumn{1}{c|}{Best}             & \multicolumn{1}{c|}{Mean}             & Var    \\ \hline
3p     & \multicolumn{1}{c|}{55.6875}          & \multicolumn{1}{c|}{\textbf{54.9837}} & 0.4401 & \multicolumn{1}{c|}{55.6582}          & \multicolumn{1}{c|}{52.6817}          & 5.0807 & \multicolumn{1}{c|}{55.6953}          & \multicolumn{1}{c|}{\textbf{54.8674}} & 0.7151 \\ \hline
3p ent & \multicolumn{1}{c|}{53.2890}          & \multicolumn{1}{c|}{53.0409}          & 0.0131 & \multicolumn{1}{c|}{53.4296}          & \multicolumn{1}{c|}{52.7154}          & 0.8765 & \multicolumn{1}{c|}{53.2968}          & \multicolumn{1}{c|}{53.0044}          & 0.0163 \\ \hline
6p     & \multicolumn{1}{c|}{55.6914}          & \multicolumn{1}{c|}{53.6941}          & 0.8542 & \multicolumn{1}{c|}{55.7792}          & \multicolumn{1}{c|}{51.9273}          & 6.0843 & \multicolumn{1}{c|}{55.7753}          & \multicolumn{1}{c|}{53.1997}          & 1.9513 \\ \hline
6p ent & \multicolumn{1}{c|}{\textbf{55.9785}} & \multicolumn{1}{c|}{54.6086}          & 0.3653 & \multicolumn{1}{c|}{\textbf{55.9179}} & \multicolumn{1}{c|}{\textbf{53.4563}} & 1.2411 & \multicolumn{1}{c|}{\textbf{55.8886}} & \multicolumn{1}{c|}{53.9241}          & 0.6900 \\ \hline
9p     & \multicolumn{1}{c|}{55.1191}          & \multicolumn{1}{c|}{51.9017}          & 2.1118 & \multicolumn{1}{c|}{55.6191}          & \multicolumn{1}{c|}{50.9688}          & 5.5388 & \multicolumn{1}{c|}{54.8867}          & \multicolumn{1}{c|}{51.6518}          & 3.4510 \\ \hline
9p ent & \multicolumn{1}{c|}{55.1093}          & \multicolumn{1}{c|}{53.6486}          & 0.2914 & \multicolumn{1}{c|}{55.5078}          & \multicolumn{1}{c|}{52.9942}          & 0.4359 & \multicolumn{1}{c|}{55.3964}          & \multicolumn{1}{c|}{53.3089}          & 0.2475 \\ \hline
\end{tabular}
\end{table}

\end{document}